\documentclass[preprint2,10pt]{aastex}

\shortauthors{Kuntz et al}
\shorttitle{OMCat}

\newcommand{\xmm}{{\it XMM-Newton}}
\newcommand{\galex}{{\it GALEX}}

\begin{document}

\title{OMCat: Catalogue of Serendipitous Sources Detected
with the {\it XMM-Newton} Optical Monitor\\
}

\author{K. D. Kuntz\altaffilmark{1,2}, 
Ilana Harrus\altaffilmark{1,2}, 
Thomas A. McGlynn\altaffilmark{2}, 
Richard F. Mushotzky\altaffilmark{2}, 
\& Steven L. Snowden\altaffilmark{2}}
\altaffiltext{1}{The Henry A. Rowland Department of Physics and Astronomy,
The Johns Hopkins University, 3701 San Martin Drive, Baltimore MD, 21218}
\altaffiltext{2}{Astrophysics Science Division Code 662, NASA/GSFC, Greenbelt, MD 20771}

\begin{abstract}
The Optical Monitor Catalogue of serendipitous sources (OMCat)
contains entries for every source 
detected in the publicly available {\it XMM-Newton} 
Optical Monitor (OM) images taken in either the imaging or ``fast'' modes.
Since the OM is coaligned and records data simultaneously 
with the X-ray telescopes on {\it XMM-Newton}, 
it typically produces images in one or more near-UV/optical bands 
for every pointing of the observatory.
As of the beginning of 2006, 
the public archive had covered roughly 0.5\% of the sky
in 2950 fields.

The OMCat is not dominated by sources previously undetected 
at other wavelengths; the bulk of objects have optical counterparts.
However, the OMCat can be used to extend optical or X-ray
spectral energy distributions for known objects into the ultraviolet,
to study at higher angular resolution objects detected with \galex ,
or to find high-Galactic-latitude objects of interest for UV spectroscopy.

\end{abstract}


\section{Overview}

\begin{deluxetable}{lr}
\tablecolumns{2}
\tabletypesize{\footnotesize}
\tablecaption{Catalogue Overview
\label{tab:over}}
\tablewidth{0pt}
\tablehead{
\colhead{} &
\colhead{} }
\startdata
Date                                 & 1 September 2006\\
Number of unique fields              & 2950 \\
Number of sources with $\sigma>3$    & 947638 \\
Number of UV sources with $\sigma>3$ & 508415 \\
Number of sources in OM unique filters\tablenotemark{a}
with $\sigma>3$ & 364741 \\
Typical positional uncertainty       & $<0\farcs45$\tablenotemark{b} \\
\enddata
\tablenotetext{a}{The OM unique filters are UVW1, UVM2, and UVW2.}
\tablenotetext{b}{As measured by the residual in position between
matched OM and USNO sources.
50\% of fields have uncertainties smaller than this value.
The distribution of uncertainties peaks at $\sim0\farcs3$.}
\end{deluxetable}

The Optical Monitor (OM) Catalogue (OMCat) contains entries 
for every point-like source
detected in imaging or fast mode OM data.
The OMCat was constructed from a complete reprocessing 
of the OM data
using the standard {\it omichain}/{\it omfchain} 
pipelines in SAS 6.5.0\footnote{
SAS was developed by members of the \xmm\ Science Survey Centre,
a consortium of ten institutions led by Prof. M. Watson
of the University of Leicester}.
For each observation (ObsID) the reprocessing created a source list;
the OMCat is a concatenation of these source lists.
Thus, if the same region of sky was observed by multiple ObsIDs,
then some sources will be listed multiple times.
Each listing should have the same coordinates
(to the limit of the astrometric accuracy)
and thus it should be reasonably obvious which listings
refer to the same source.
We have opted to retain multiple listings
(rather than to combine them into a ``mean'' entry) 
to retain any useful information of temporal variability
in an easily accessible manner.
An overview of the catalogue statistics is given in Table~\ref{tab:over}.

In this paper
\S2 provides a brief description of the OM and its primary observation modes,
\S3 describes the standard pipeline processing,
the further processing done to produce the source lists,
and the output products that are unique to our processing.
\S4 contains some useful statistics describing the catalogue.
In \S5 we demonstrate the extent to which the OM filter set
allows photometric classification of point-like sources,
and in \S6 we explore some of the scientific uses of the OM-specific bands. 

\section{Brief Description of the OM}

{\bf The telescope and detector:}
The OM \citep{mea2001} is a 30 cm f/12.7 Ritchey Chretien telescope
coaligned with the X-ray telescopes
and operating simultaneously with them.
The detector is a micro-channel plate intensified 
charge-coupled device (CCD).
Photons striking a photocathode
produce electrons that are amplified 
by two successive micro-channel plates.
The electron clouds then strike a phosphor,
and the resulting photon splashes are recorded by a CCD;
the location of the photon splash is centroided on board.
The centroids are stored in units of 1/8 of a CCD pixel.
Since it is these photon splashes that are recorded by the CCD,
rather than individual photons,
the CCD is read out very rapidly (every 11$\mu$s),
and the centroids of the photon splashes determined and stored.
Thus, the CCD is used more like a photon-counting device than an accumulator,
although it is an image, rather than an event list, that is produced.
The photocathode is optimized for the blue and ultraviolet.
The ``native'' pixel size is $0\farcs476513$
and the point spread function (PSF) FWHM is $1\farcs4$-$2\farcs0$ 
depending upon filter.
The largest possible field of view (FOV) is roughly $17\arcmin\times17\arcmin$.

\begin{figure}[h!]
\epsscale{1.0}
\plotone{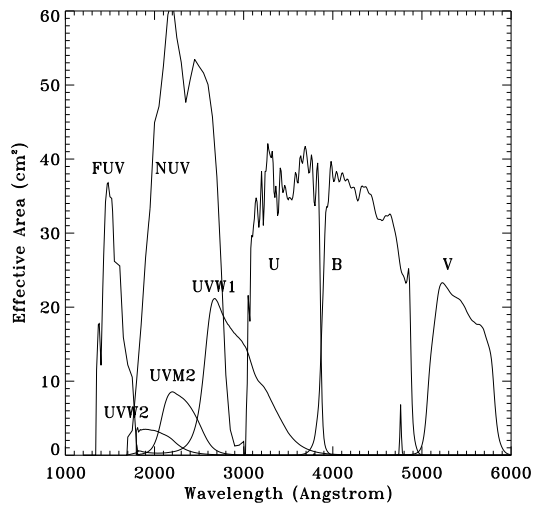}
\caption{\footnotesize The OM and {\it GALEX} filter sets.
The two lowest wavelength filters are
the {\it GALEX} FUV and NUV filters.
The next three are the OM UVW2, UVM2, and UVW1 filters.
The three highest wavelength filters are the OM U, B, and V filters.
\label{fig:filts}}
\end{figure}

\begin{deluxetable}{lrrrrr}
\tablecolumns{6}
\tabletypesize{\footnotesize}
\tablecaption{OM Filters
\label{tab:filters}}
\tablewidth{0pt}
\tablehead{
\colhead{Name} &
\colhead{$\lambda_0$\tablenotemark{a}} &
\colhead{$\lambda_{max}$\tablenotemark{b}} &
\colhead{FWHM} &
\colhead{PSF FWHM} &
\colhead{Peak} \\
\colhead{ } &
\colhead{(\AA)} &
\colhead{(\AA)} &
\colhead{(\AA)} &
\colhead{(arcsec)} &
\colhead{Mag.} }
\startdata
V    & 5407 & 5230 & 684 & 1.35 & 19.0\\
B    & 4334 & 3980 & 976 & 1.39 & 19.7\\
U    & 3472 & 3270 & 810 & 1.55 & 19.5\\
UVW1 & 2905 & 2680 & 620 & 2.0  & 19.3\\
UVM2 & 2298 & 2210 & 439 & 1.8  & 18.3\\
UVW2 & 2070 & 2000 & 500 & 1.98 & 17.6\\
WHITE\tablenotemark{c}&      &      &     & & 22.2\\
\enddata
\tablenotetext{a}{Effective wavelength}
\tablenotetext{b}{Wavelength of maximum transmission}
\tablenotetext{c}{An ``open'' filter}
\end{deluxetable}

The OM has a smaller FOV than \galex\
(a 1.2$\arcdeg$ circle)
but better angular resolution
\citep[\galex\ has a $4\farcs5$ FWHM PSF in its FUV filter (1350-1750 \AA , 
and a $6\farcs0$ FWHM PSF in its NUV filter (1750-2800 \AA)][]{galex_morrissey}.
The effective areas of the OM and \galex\ filters are shown
in Figure~\ref{fig:filts} 
while the OM filter particulars are given in Table~\ref{tab:filters}.
Thus, the OMCat data in the UVW2 and UVM2 filters
provide an excellent higher resolution complement to the \galex\ NUV data,
while UVW1 data is somewhat redder than the \galex\ band.
The {\it Swift} UVOT is, essentially, an improved OM, 
with similar filters,
so comparison of data in this catalogue with UVOT data
should be straightforward.

{\bf The observation modes:}
Due to the onboard centroiding, memory limitations, and telemetry limitations,
setting the OM observation mode has to be a balance 
of temporal resolution and spatial coverage;
the higher the temporal resolution the lower the spatial coverage.
As a result, the OM allows a large number of observing modes
that place different emphases on temporal and spatial optimization.
These modes define different ``science windows'' 
covering only portions of the entire FOV; 
events falling outside of those windows are discarded.
There are two primary observation modes at the extremes:
the default ``imaging'' mode and the default ``fast'' mode.

\begin{figure*}
\epsscale{1.0}
\centerline{\plotone{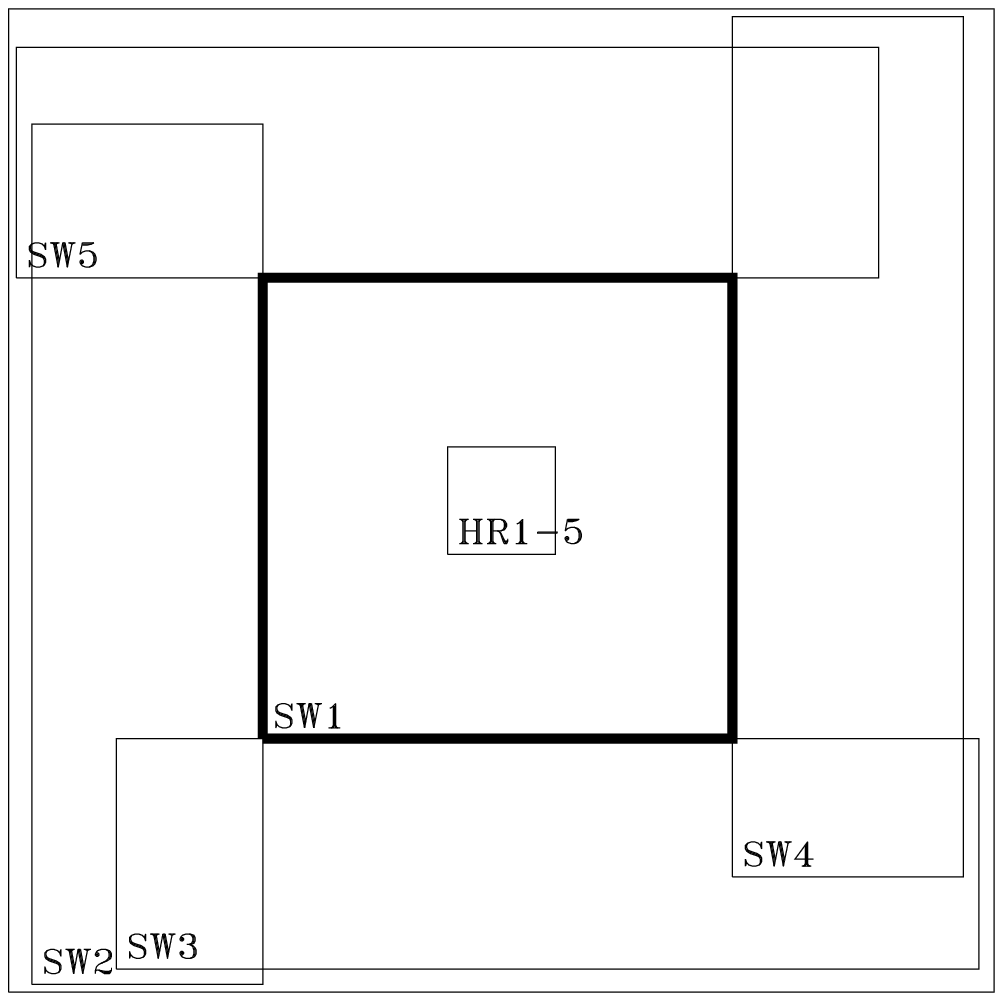}\plotone{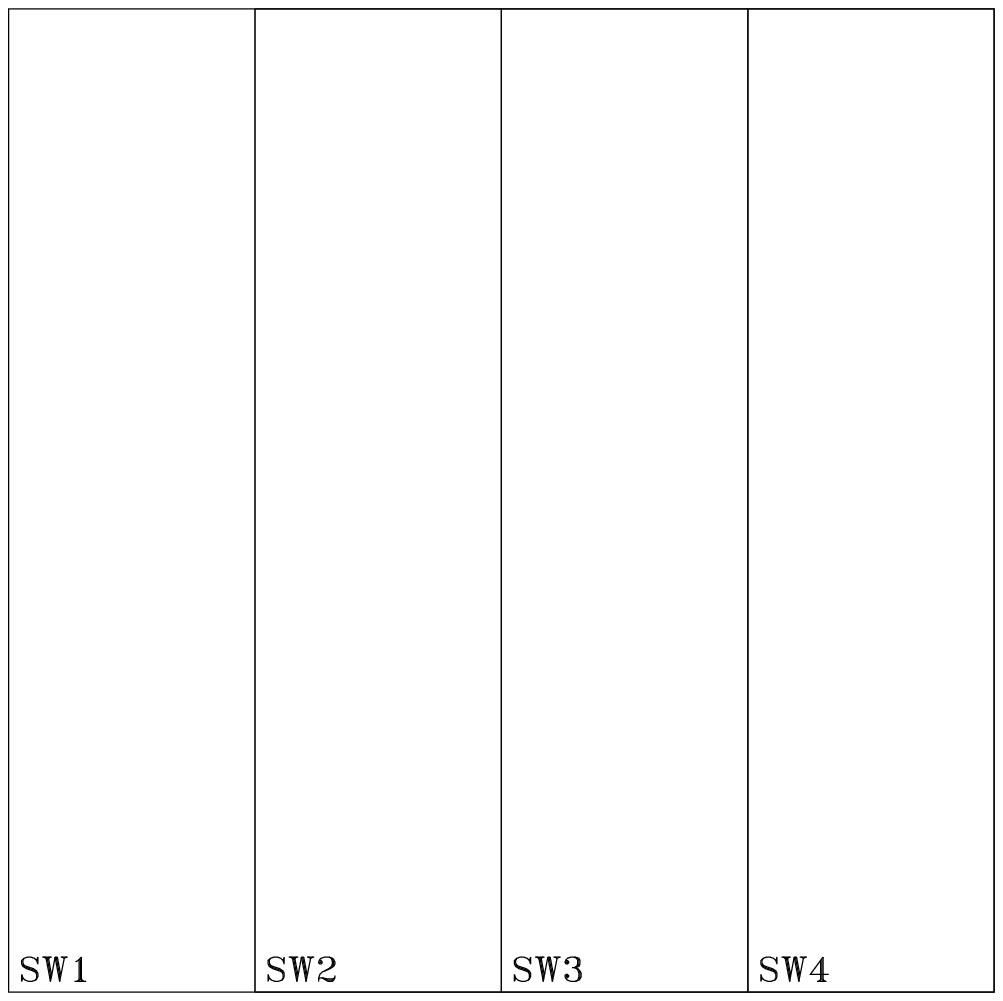}}
\caption{\footnotesize {\bf Left:} The FOV for the default mode.
The plot has dimension of 2048$\times$2048 unbinned pixels
or $16\farcm26\times16\farcm26$.
{\bf Right:} The FOV for the ENG-2 and ENG-4 modes.
The plot has the same dimensions as the previous panel.
\label{fig:fov}}
\end{figure*}

The default imaging mode consists of five consecutive sub-exposures,
each of which employs two science windows; 
one high-resolution window and one low-resolution window.
Note that resolution, in this case, 
refers to the degree to which the image is sampled,
not to an intrinsic change in the PSF size.
The FOV of the default imaging mode is shown 
in the left-hand panel of Figure~\ref{fig:fov}.
The high-resolution window 
(roughly $5\arcmin\times5\arcmin$,
marked ``HR'' in the figure)
is always located at the center of the FOV.
The five low-resolution windows 
(marked ``SW'' in the figure)
cumulatively cover the entire FOV
(roughly $17\arcmin\times17\arcmin$)
with a center square surrounded by five rectangular regions.
For any number of reasons, 
not all of the sub-exposures of a default image may actually be taken,
but, for the default imaging mode,
there will always be a high-resolution sub-exposure
for each low-resolution sub-exposure.
It should be noted that multiple observing modes
may be used during the course of a single observation,
so not all exposures of a single observation need cover the same region.
The use of five different science windows to cover the FOV,
with some overlap between the windows,
means that the exposure is not uniform across the FOV.

There are two other common full-field low-resolution modes,
``ENG-2'' and ``ENG-4'',
which are also included in our processing.
The FOV for these modes are shown 
in the right-hand panel of Figure~\ref{fig:fov};
the two modes cover the same area but with different binning,
the ``ENG-2'' mode having twice the pixel size as the ``ENG-4'' mode.
There is no repeated high-resolution window for these modes.

The default fast mode uses the same windows as the default imaging mode
with the addition of a third science window 
(roughly $10\farcs5\times10\farcs5$) at an observer-defined location
(typically the center of the FOV).

{\bf The default modes:}
If the observer did not specify an OM mode,
and there was no bright source in the FOV,
the OM took exposures in the default imaging mode.
For the first two years of the mission
the default filters were B, UVW2, U, and UVW1, in order of priority.
The filter priority was then changed to UVM2, UVW1, and U,
in order to optimize the use of the unique capabilities of the OM.

{\bf Magnitude system:}
The standard OM processing produces {\it instrumental} magnitudes.
We have opted to continue to work in the instrumental magnitude system.
The definition of the instrumental magnitudes 
and the current conversion to AB magnitudes is given in the \xmm\ 
User's Handbook\footnote{http://heasarc.gsfc.nasa.gov/docs/xmm/uhb/XMM\_UHB.html}
In rough terms,
the V and B the instrumental magnitudes are similar to AB magnitudes,
but the UV the instrumental magnitudes are typically smaller
than the AB magnitudes by a magnitude to a magnitude and a half.

\section{Processing}

{\bf Image mode processing:}
For the most part, we have used the standard {\it omichain}\footnote{
SAS routines are documented at
http://xmm.vilspa.esa.es/sas/6.5.0/doc/packages.All.html}
processing with the default settings; exceptions are detailed below.
The standard {\it omichain} processing (processing for images)
handles the images produced by each science window separately.
For each science window image {\it omichain}
applies a flatfield.
The photon splash centroiding algorithm calculates the centroid
to 1/8 of a pixel, but due to the algorithm,
not all values are equally likely.
This problem results in ``modulo-8'' fixed pattern noise.
The {\it omichain} processing applies a redistribution 
to correct for this effect.

For every science window image
{\it omichain} runs a source detection algorithm,
measures the count rates for the sources,
and applies a calibration to convert to instrumental magnitudes.
Once all of the science windows are processed,
{\it omichain} produces a ``master'' source list by combining
the source list for each science window,
matching sources in common between the lists,
and determining the mean $(\alpha,\delta)$ for each source.
The standard processing has the option to use an external catalogue
to correct the coordinates of the master source list;
we have used this option with the USNO-B1 catalogue \citep{usno}.
It should be noted that
the coordinate correction using the USNO-B1 catalogue
will fail if there are too few matching sources,
in which case no significant solution can be found.
The {\it omichain} algorithm requires at least ten matches
in order to produce a significant coordinate correction.
If there are too many sources in the field,
coordinate correction will also fail,
presumably because some fraction of matches are spurious
and the solution will not converge.
We have found that the coordinate correction
{\it can} fail for almost any density of sources,
though we did not determine the cause of that failure.

In addition to combining the source lists for all the filters,
{\it omichain} mosaics the low-resolution science window images
(but not the high-resolution science window images) for each filter.
The world coordinate system (WCS) of the first science window 
for a given filter sets the coordinate system of the entire image mosaic.
Note that the standard {\it omichain} processing can correct
the master source list, but not the images.
Further, the correction using an external catalogue
will be applied only if there are at least ten sources.
One does have the option of doing the same correction
to the individual science windows, but there are often
not enough sources in a single science window to perform such a correction.
We have applied the correction derived from the external catalogue 
to the low-resolution mosaics.
Since the mosaic images use the WCS 
from the first science window in the mosaic, 
we compared the $(\alpha,\delta)$ in the source list
for the first science window
to the $(\alpha,\delta)$ in the USNO corrected master list.
We then determined the $(\Delta\alpha,\Delta\delta)$
which must be added to the $(\alpha,\delta)$ of the first source list
in order to obtain the $(\alpha,\delta)$ in the USNO corrected master list.
We then applied that correction to the 
header keywords of the mosaicked image.
It must be noted that
since the telescope can drift by 1-2 arcseconds between exposures,
this correction is done separately for each filter.

\begin{figure*}
\epsscale{1.0}
\plotone{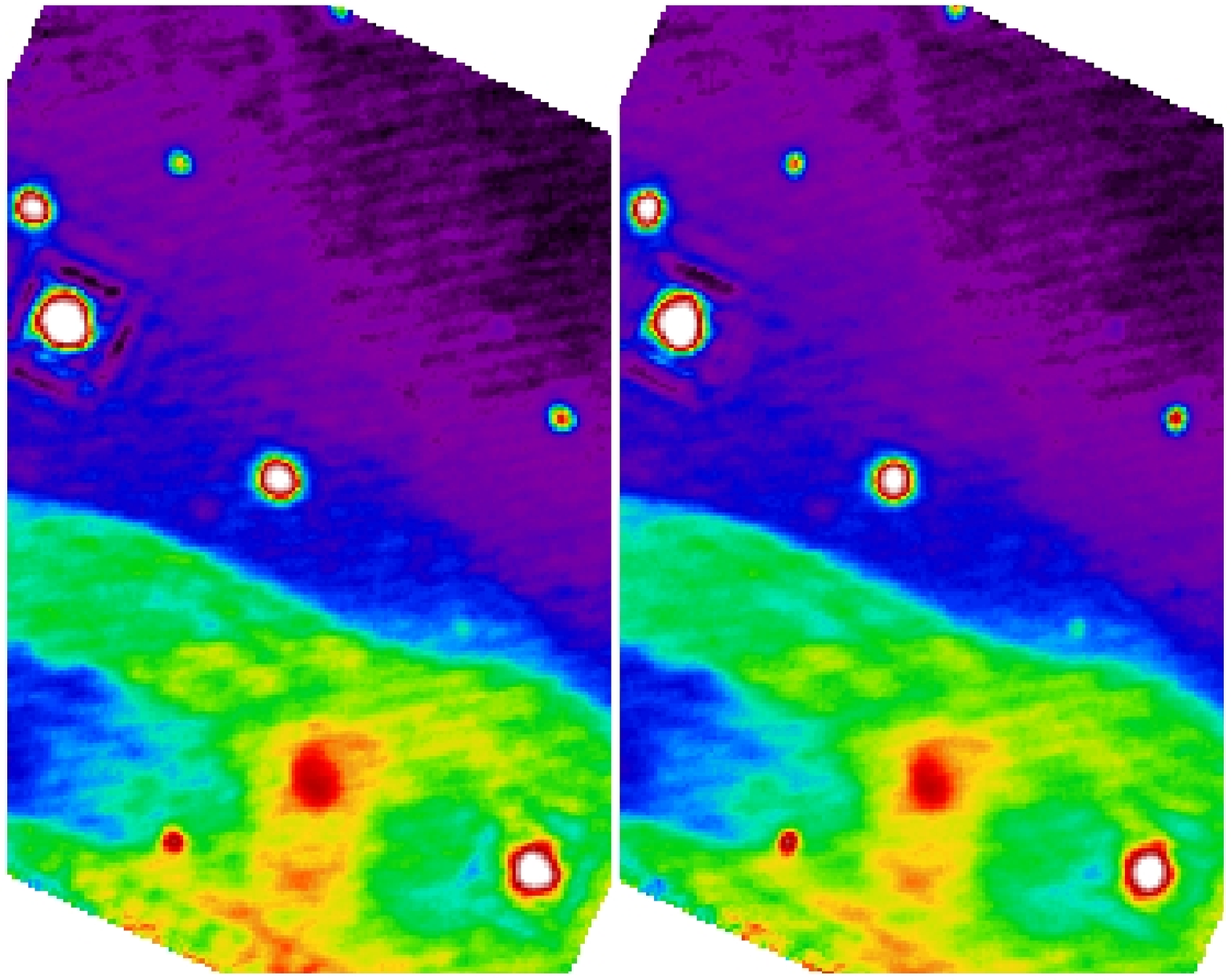}
\epsscale{1.0}
\plotone{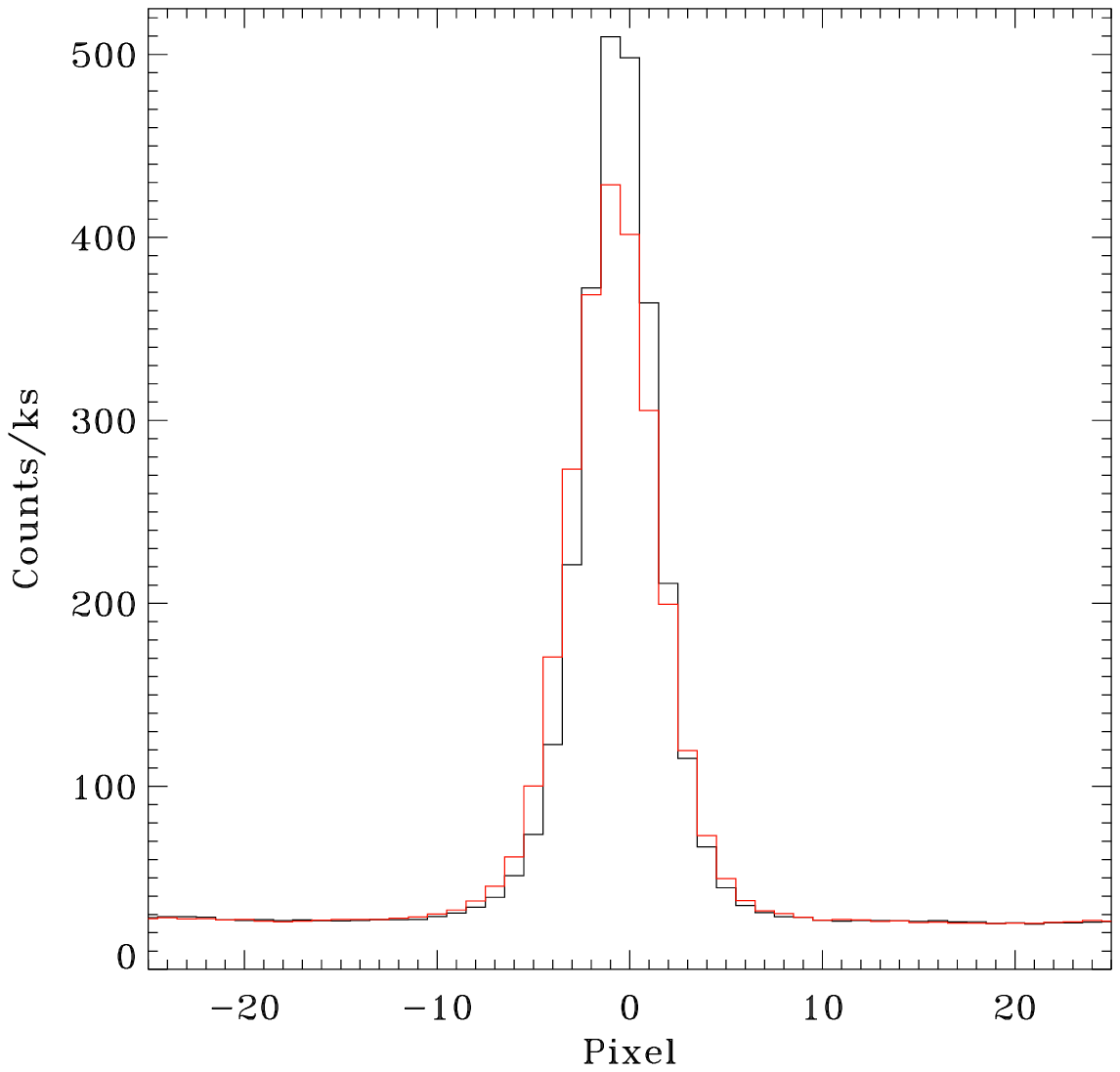}
\caption{\footnotesize {\bf Top:} Image created from 25 high-resolution
science windows using {\it ommosaic} (left panel) compared to the same
image created with our processing (right panel).
Note that the central point source is rounder after our processing.
{\bf Bottom:} Comparison of the profile of the central point source
{\it red:} with {\it ommosaic} alone {\it black:} with our processing.
The FWHM improves by almost a pixel in this case
and the peak intensity increases by 15\%.
\label{fig:hires_demo}}
\end{figure*}

The standard {\it omichain} processing does not combine
the images from the high-resolution science windows.
However, we have done so,
though not with {\it ommosaic}, the standard SAS tool.
The {\it ommosaic} program uses the WCS keywords to determine
the offsets needed to align the WCS frames
of the individual science windows before summing.
We allow the WCS of the summed image to be set 
by the first science window.
For each successive high resolution image,
we compare the source list to the source list from the first image,
determine the $(\Delta\alpha,\Delta\delta)$ required to match the source
lists, and apply that offset to the image before adding it to the mosaic.
Not all sources are used for determining the offsets,
only sources appearing in at least half of the images;
this selection removes sources with poorly determined positions.
The offsets are rounded to the nearest integer pixel;
subpixelization did not seem to produce a significant improvement
in the resultant PSF, and so was not used for this processing.
Compared to the direct sum of the images made by {\it ommosaic},
our processing does improve the PSF of the summed image,
sometimes improving the FWHM by as much as a pixel
(see Figure~\ref{fig:hires_demo}).
We compare the source list from the first image
with the master source list to correct the summed image
in the same manner used for the low-resolution mosaics.

\begin{figure}
\epsscale{1.0}
\plotone{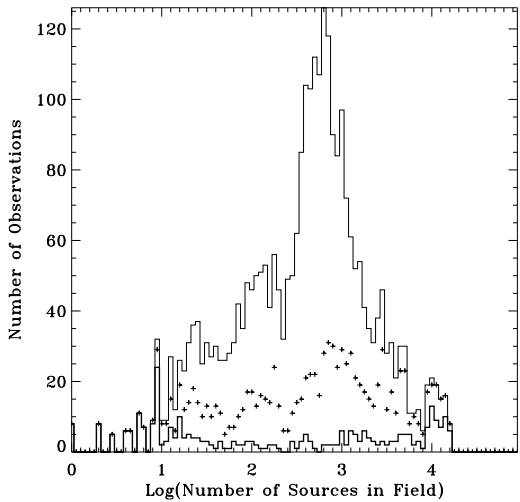}
\caption{\footnotesize Histogram of the number of observations 
as a function of source density.
The thin solid line is the total number of observations.
The crosses show the number of observations for which the pipeline
coordinate correction (i.e., that done by the SAS {\it omichain} task)
failed as a function of the number of sources.
The thick line shows the number of observations for which the post-pipeline
coordinate correction (our software) failed as a function of the number of sources.
\label{fig:bad_corr}}
\end{figure}

{\bf Further coordinate correction:}
After our initial processing of the public archive
we found that the pipeline coordinate correction (done by {\it omichain})
using the USNO catalogue failed for $\sim38$\% of the fields.
Further, the failure was not limited to extremely high
or extremely low source densities (see Figure~\ref{fig:bad_corr})\footnote{
We have not had the opportunity to trace the root of this problem.
However, we note that a disproportionate number of ObsIDs
lacking pipeline corrections {\it seemed} 
to have a single science window that was strongly discrepant from the others.}.
We thus found it worthwhile to create our own coordinate
correction routine (the ``post-pipeline'' correction) using the USNO catalogue.
Although the bulk of fields need only a small correction,
some fields need substantial corrections ($\sim2\arcsec$).
Thus, although we attempt to find high precision
corrections for all fields, 
it is still worthwhile to find lower accuracy corrections 
for those fields that do not have a large enough number of matches
with the USNO catalogue to attempt a high precision solution.

We used a fairly simple and robust algorithm 
for matching the OM sources to the USNO sources.
If there were $>10$ matches we iteratively solved
for the offset in $(\alpha,\delta)$
that minimized the offset between the OM source list with the USNO source list.
By iterating the solution we could eliminate some portion of the false matches.
We have not solved for a rotation for two reasons:
1) adding a rotation to the fit did not significantly improve
the solution, and
2) given that there are systematic offsets from one science window to another,
the rotation could be strongly biased by the offset of a single science window.
If there were $3<n<10$ matches we merely calculated
the mean offset between the OM and USNO sources,
and used that offset as the coordinate correction.
If there were $<3$ sources we did not attempt a correction.
We applied the same correction to the individual images 
that we applied to the source lists.

For each OM source in the source list with
the {\it omichain}-calculated maximum-liklihood significance in any filter $>3$,
the matching algorithm finds the closest USNO source.
It then creates the distribution of the distances
between the OM sources and their closest USNO counterparts.
If all of the OM sources had USNO counterparts,
then this distribution would be a Gaussian
whose width is the coordinate uncertainties of the two catalogues
and whose peak is the offset between the two catalogues.
If there were no true matches between the OM sources
and the USNO catalogue,
then the distribution would be given roughly by
the probability distribution for the minimum distance
between a given point and a uniform distribution of sources:
\begin{equation}
P(r)=e^{\left[-\rho\pi r^2\right]} \rho\pi[(r+\delta r)^2-r^2] e^{\left[-\rho\pi [(r+\delta r)^2-r^2]\right]}
\end{equation}
where $\rho$ is the surface density of sources and
$\delta r$ is the binsize of one's histogram of distances.
For this distribution both the peak of the distribution
and the width of the distribution scale as $\rho^{-0.5}$.
We expect that some fraction of the OM sources have true USNO matches
and that the remainder will not.
As a result, the observed distribution of sources
has a sharp peak with a width of $\sim0.3\arcsec$ due to matches
and a low, broad distribution for the spurious matches.
This algorithm has problems with high density regions;
for source densities of 5000 sources/image
(0.005 sources arcsec$^{-2}$) 
the distribution of spurious matches
peaks at $6.3\arcsec$ with the lower half-maximum at $2\arcsec$.
Although the peak of the matching sources typically has $r\lesssim3\arcsec$,
the true match rate is likely to be small compared to the spurious match rate,
and so it is difficult, if not impossible,
to find the true match peak in this distribution.
However, at these source densities, source confusion is a serious problem
as well, so even if the matching algorithm worked, 
the coordinate solution would remain problematic.

For the matching algorithm,
we simply fit the distribution with a Gaussian.
If the width of the Gaussian is smaller than $0\farcs7$,
then the algorithm takes all of the sources within $3\sigma$
of the peak of the distribution as real matches.
An initial solution is determined from those matches,
and a fit is made in the image coordinate frame
to find the $(\alpha,\delta)$ offset that minimizes
the distance between the OM sources and their USNO matches.
The source matching is redone with the new offset,
and the process is iterated until it converges.
After application of our coordinate correction routines,
only $\sim 14$\% of the fields remained without any coordinate correction.
Besides fields with very few objects,
the fields without coordinate corrections were characterized
by very broad distribution of the matches suggesting
a combination of large pointing error and large source density,
and thus a large number of spurious identifications.

For fields where the pipeline processing found a good coordinate solution
using the USNO catalogue our coordinate correction 
was not significantly different.
However, our coordinate correction did improve the mean distance 
between OM and USNO sources for $\sim44$\% of fields,
and provided the only coordinate corrections for $\sim23$\% of fields.
(Figure~\ref{fig:bad_corr})

The RMS residual between the OM sources and the matching USNO sources 
was calculated for every field.
A value of zero indicates that there was no coordinate solution.
Coordinate solutions with RMS residuals $>0\farcs6$
should be considered to be poor.

{\bf Fast mode processing:}
The bulk of the fast mode processing is concerned 
with the production of light-curves of the source.
The fast mode images consist of 
$10\farcs5\times10\farcs5$ regions containing,
typically, a single source.
We combine all of the images for each filter
using the same method applied to the high-resolution images.

{\bf Further processing:}
Further processing is required to provide a more useful source list 
to be incorporated into the OMCat.
To the standard image catalogue (a binary fits table)
we add images of each source from each filter.
Each ``postage stamp'' image is $19\times19$ pixels in size,
extracted from the low-resolution image mosaics
(the pixel size is $0\farcs95$ 
and the image is $18\farcs1\times18\farcs1$ in size).
Since the sources were derived from all of the science windows,
some sources can fall in high-resolution science windows
without low-resolution counterparts.
In that case the postage stamp is extracted from the high-resolution image
and binned to the same resolution and size
as the other postage stamps.
Sources that appear only in ``fast'' science windows are treated similarly.
Note that postage stamps are extracted from all of the available filters,
not just the filters for which the source was detected;
many postage stamps may thus appear to be empty.

{\bf Processing summary:}
For each ObsID our processing produces
a coordinate corrected source list,
a coordinate corrected low-resolution mosaicked image for each filter,
a coordinate corrected high-resolution mosaicked image for each filter
(if possible), or
a summed fast mode image.

{\bf Caveats:}
1) Individual science windows may be significantly offset (1-2$\arcsec$)
from the remainder of the mosaic.
In this case the correction by use of the USNO-B1 catalogue
will not be wholly satisfactory, 
and sources will appear to be offset in the postage stamps.
The extent of this problem for any individual source can
determined by looking at the ``RMS\_RESID'' column which contains
the RMS residual from the fit of source list to the USNO-B1 catalogue.

2) The source lists will contain spurious sources,
sources due to ghost images, diffraction spikes, readout streaks,
saturation around bright sources, and other effects.
Some of these sources can be removed 
by consulting the data quality flag (Q\_FLAG parameter)
Similarly, confused sources are flagged by the C\_FLAG parameter.
However, we have found that filtering out sources
with signal-to-noise ratios of less than three was a more efficient means
of removing spurious sources than reference to the quality flags.

\begin{figure}[h]
\epsscale{1.0}
\plotone{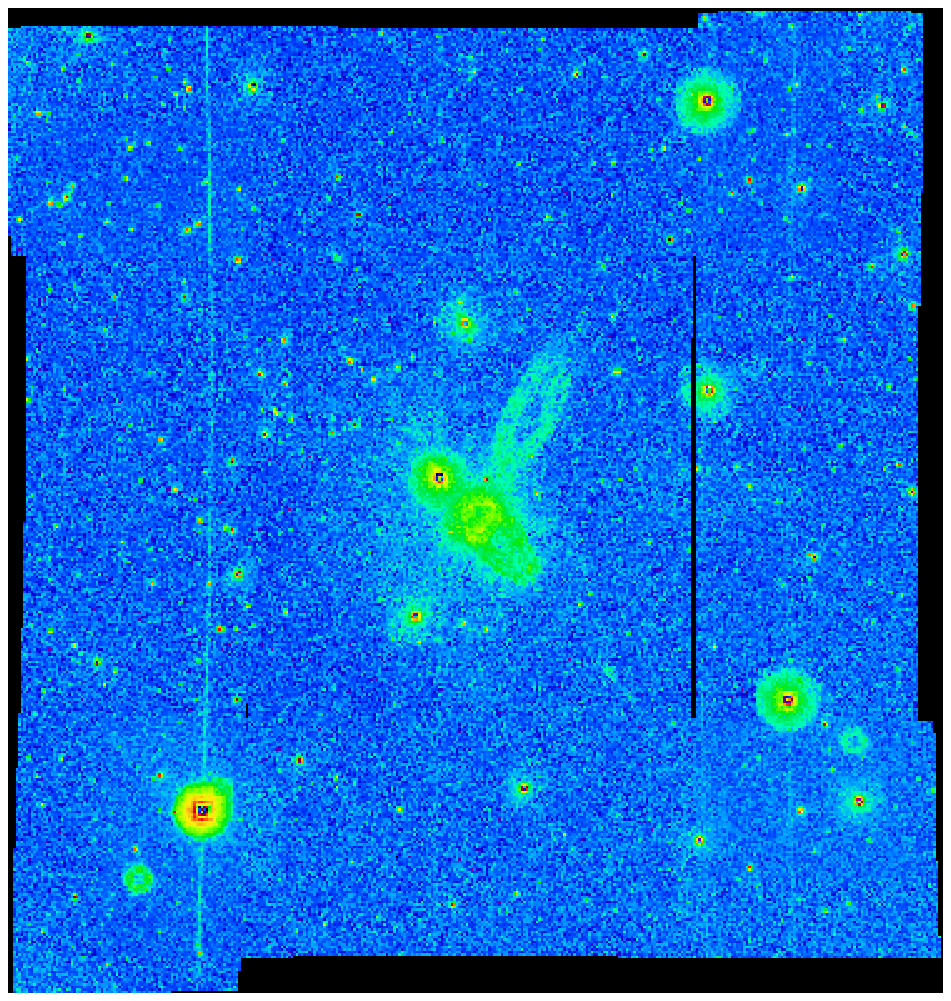}
\caption{\footnotesize An image displaying typical image problems:
the UVW1 image from obsid 0000110101.
Note the readout streaks, ghosts, and diffraction spikes.
Note also that the individual science windows do not always
overlap (or butt) correctly,
as shown by the black vertical stripe in the right hand part of the image.
\label{fig:demo_imag}}
\end{figure}

\begin{figure*}
\includegraphics[scale=0.85,angle=90]{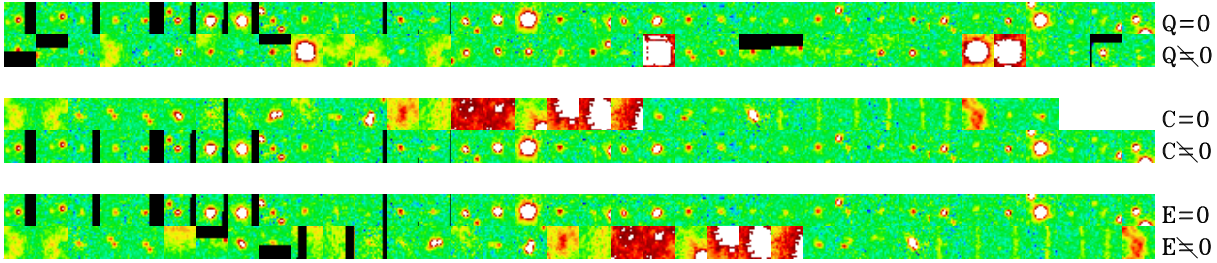}
\caption{\footnotesize UVW1 filter images of individual sources from obsid 0000110101.
The first (top) row of images contains UVW1 band images of the first 36 sources
in the source list which have Q\_FLAG$=0$, or good quality sources.
The second row of images contains UVW1 band images of the first 36 sources
in the source list which have Q\_FLAG$\ne0$, or poor quality sources.
The third row of images contains UVW1 band images of the 33
sources with C\_FLAG$=0$ (unconfused).
The fourth row of images contains UVW1 band images of the first 48 sources
in the source list which have C\_FLAG$\ne0$ (confused).
The fifth row of images contains UVW1 band images of the first 48 sources
in the source list with E\_FLAG$=0$ (non-extended).
The sixth (bottom) row of images contains UVW1 band images of the 38 sources
with E\_FLAG$\ne0$ (extended).
\label{fig:demo_source}}
\end{figure*}

\begin{figure*}
\includegraphics[scale=0.85,angle=90]{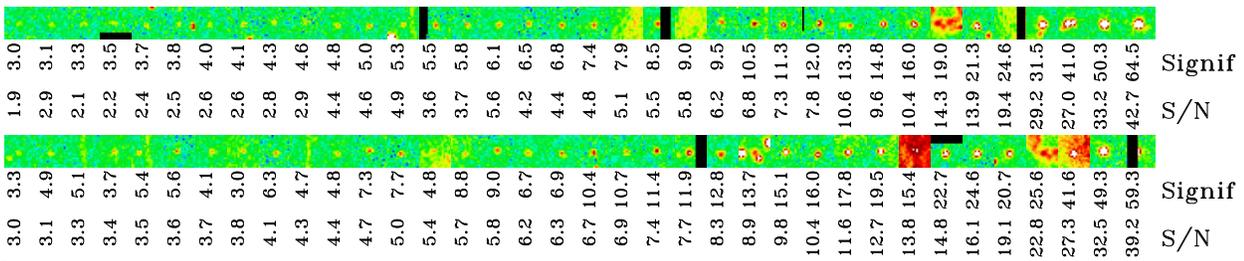}
\caption{\footnotesize The top row of images shows the effect of increasing significance,
as characterized by the SIGNIF parameter in the UVW1 band.
The images are some of the 2143 sources from obsid 0000110101.
For this row of images, we sorted the sources by SIGNIF,
and sampled every $(2143/36)^{th}$ source to represent
the entire range of the SIGNIF parameter.
The value of the SIGNIF parameter is given
directly to the below the images,
and the source signal-to-noise ratio directly below that.
Similarly, the bottom row of images samples sources
of increasing value of the source signal-to-noise ratio;
the values of the SIGNIF parameter and the signal-to-noise ratio
are shown below the images.
\label{fig:demo_signif}}
\end{figure*}

Figure~\ref{fig:demo_imag} and Figure~\ref{fig:demo_source}
demonstrate the extent to which the data quality, confusion,
and extension flags can be used.
Figure~\ref{fig:demo_imag} shows an image with typical difficulties:
ghosts, readout streaks, and diffraction spikes.
Figure~\ref{fig:demo_source} shows images of some of 
the individual sources from the image in Figure~\ref{fig:demo_imag}.
The first row of images contains UVW1 band images of the first 36 sources
in the source list which have Q\_FLAG$=0$, or good quality sources.
The second row of images contains UVW1 band images of the first 36 sources
in the source list which have Q\_FLAG$\ne0$, or poor quality sources.
It is clear that some of the ``good quality sources'' are unreliable,
while some of the ``poor quality sources'' are reliable.
The third and fourth rows of images contain, respectively,
sources with C\_FLAG$=0$ (unconfused) and C\_FLAG$\ne0$ (confused);
again the sources are taken in the order of the source list
(which happens to be from low R.A. to high R.A.)
and all images are in the UVW1 band.
As can be seen, the confusion flag is not reliable.
The fifth and sixth rows of images contain, respectively,
sources with E\_FLAG$=0$ (non-extended) and E\_FLAG$\ne0$ (extended);
this flag does seem to be robust.
Figure~\ref{fig:demo_signif} shows images of some 
of the individual sources from the image in Figure~\ref{fig:demo_imag},
selected to demonstrate the effect of the SIGNIF parameter.
The top row of images shows the effect of increasing significance,
as characterized by the SIGNIF parameter in the UVW1 band.
For this column of images, we sorted the sources by SIGNIF,
and sampled the sequence uniformly to represent
the entire range of the SIGNIF parameter.
The value of the SIGNIF parameter is given 
directly below the images,
and the source signal-to-noise ratio directly below that.
Similarly, the bottom row of images samples sources
of increasing value of the source signal-to-noise ratio;
the values of the SIGNIF parameter and the signal-to-noise ratio
are shown below the images.
Neither the SIGNIF parameter nor the signal-to-noise ratio
provides a completely reliable estimator of the reality
of a particular source.

3) Although we provide the mosaicked low-resolution and
mosaicked high-resolution images through the archive,
these images, according to the SAS documentation,
should not be used for photometry.
There are a number of corrections in the photometric reduction
which can not be made from the mosaicked images.
However, since coincidence-loss and dead-time corrections are small
for faint sources and faint extended emission, 
photometry of faint sources using standard non-SAS tools
is possible from the mosaics,
as has been demonstrated by D. Hammer (private communication).

4) Although there may be substantial exposure for a given mosaic,
the detection limit in the current catalogue is not that of the mosaic
since the source detection is done on the individual science windows
rather than on the mosaicked images.
Since the individual science windows have a median exposure of $\sim1800$ s,
and the bulk of the individual science windows have exposures of 1000 s,
the detection limit of the OMCat is more uniform but somewhat lower
than one would expect from the mosaic exposure times.

{\bf Availability:}
The bulk of the data in the OMCat can be accessed 
either through the Browse facility at the 
High Energy Astrophysics Science Archive (HEASARC)
or through the Multimission Archive at STScI (MAST).
Slightly more information for each source
(shape, confusion flags, and quality flags, as a function of filter)
are retained in the source lists for individual ObsIDs,
and can be downloaded from the HEASARC with the rest of the OM data.

\section{OMCat Statistics}

\begin{figure}
\epsscale{1.0}
\plotone{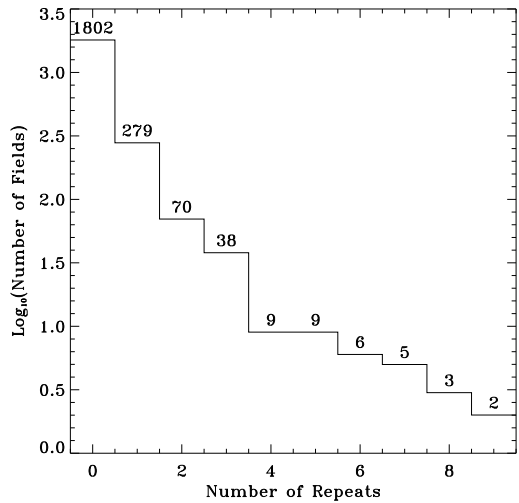}
\caption{\footnotesize The histogram of the number of fields
as a function of the number of times
that the observation was repeated.
Observations were considered to be repeated if
the pointing directions between two observations
were offset by no more than half an arcminute.
Zero repeats indicates a field with only one observation.
The numbers give the number of fields
for each number of repetitions.
\label{fig:dup_stat}}
\end{figure}

{\it Observation Statistics:}
Of the 4373 observations that were public by 1 September 2006,
2950 observations had OM imaging mode data 
and 202 had OM fast mode data.
About 25\% of the fields imaged
were observed more than once,
allowing some measure of temporal variability.
However, the number of fields with multiple observations
is a very strongly declining function
of the number of repetitions (see Figure~\ref{fig:dup_stat}).

\begin{figure}
\epsscale{1.0}
\plotone{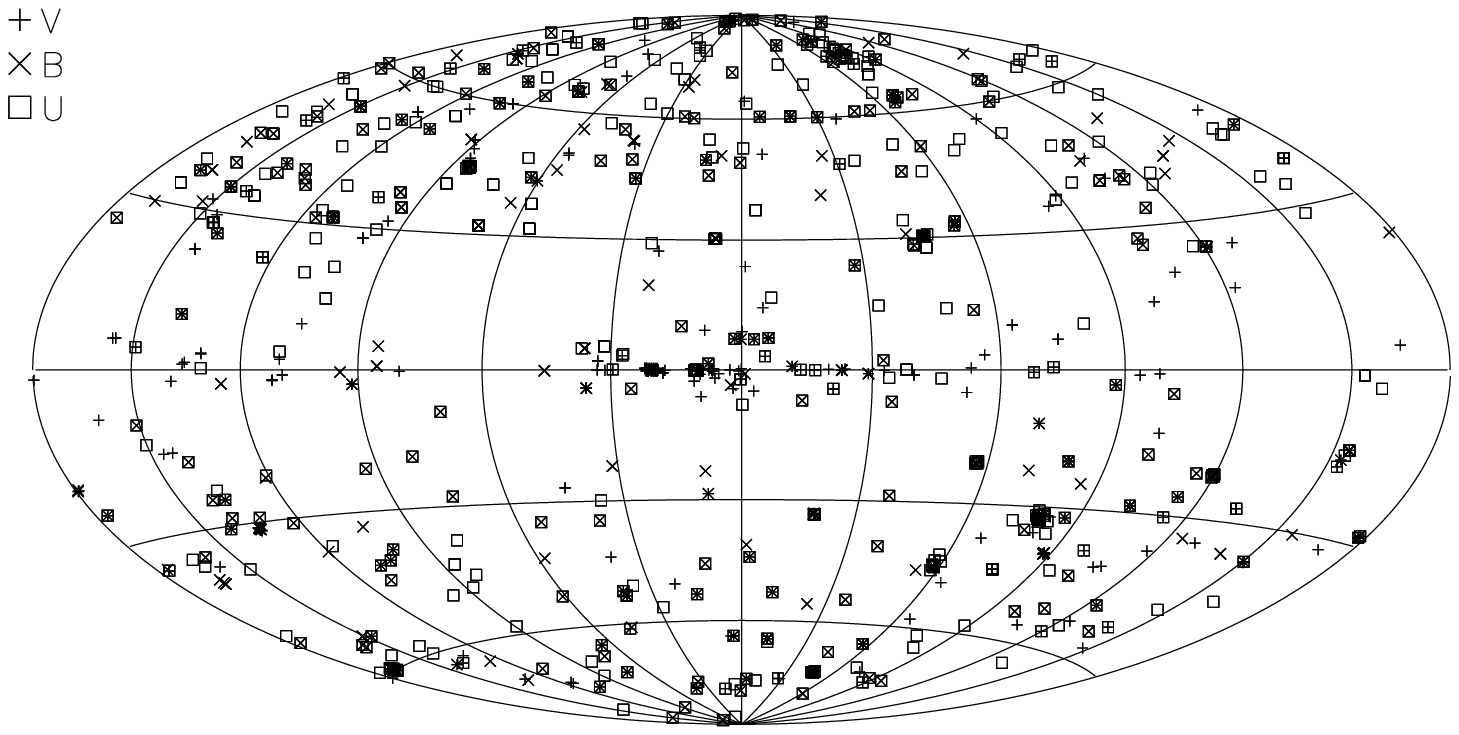}
\epsscale{1.0}
\plotone{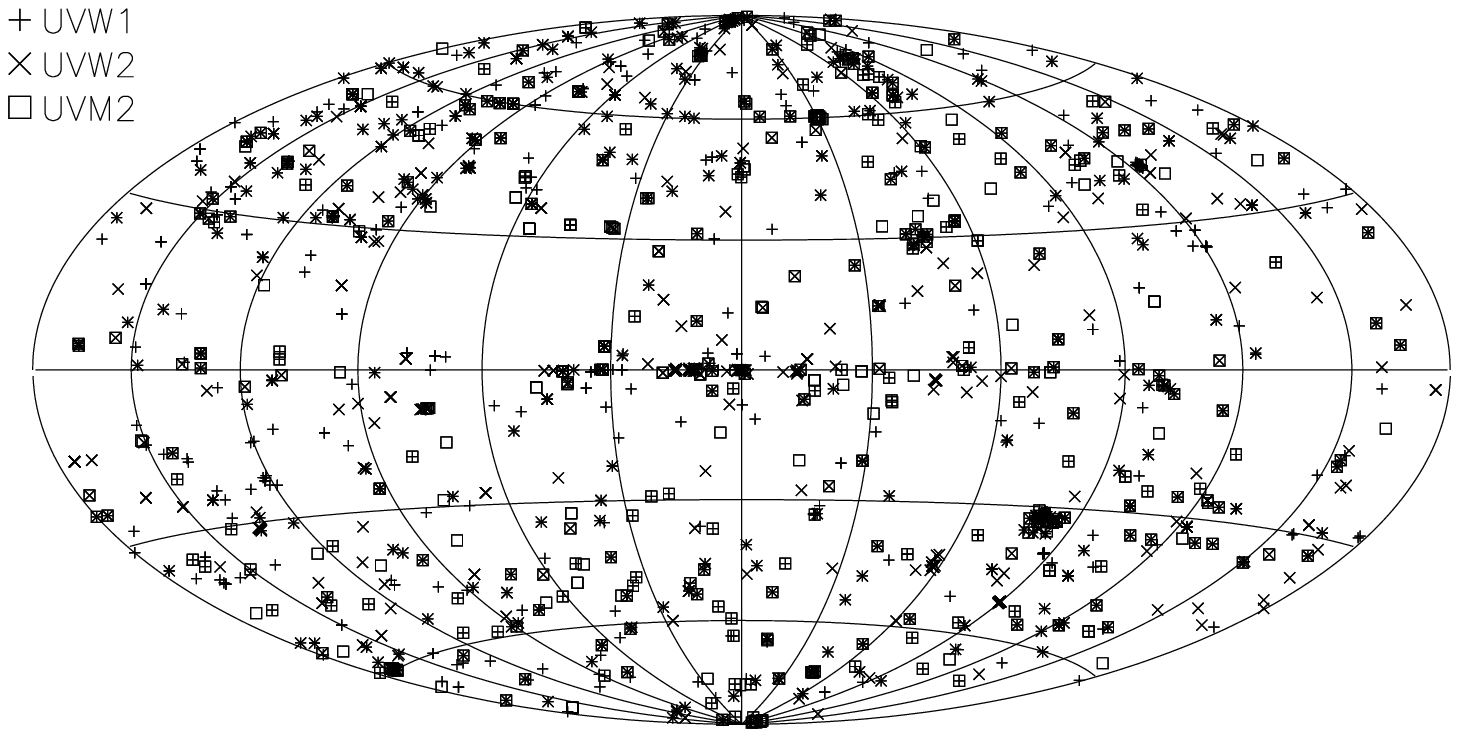}
\caption{\footnotesize The distribution of observations over the sky.
The Aitoff coordinate system is centered on $(\ell,b)=(0\arcdeg,0\arcdeg)$
with positive longitudes towards the left.
\label{fig:sky_map}}
\end{figure}

As can be seen in Figure~\ref{fig:sky_map},
the Galactic plane has a high density of observations
(particularly towards the Galactic center),
the region with $|b|<30\arcdeg$ has a lower density of observations,
and the region with $|b|>30\arcdeg$ is relatively uniform,
though Coma and the Magellanic clouds 
have visible concentrations of observations.
The imaging mode observations cover a cumulative $\sim0.5$\% of the sky.

\begin{figure}
\epsscale{1.0}
\plotone{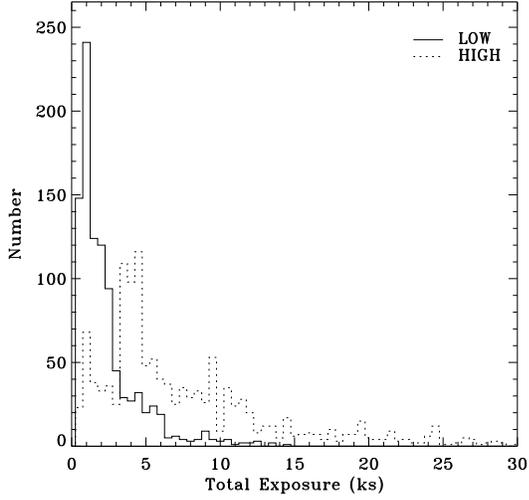}
\epsscale{1.0}
\plotone{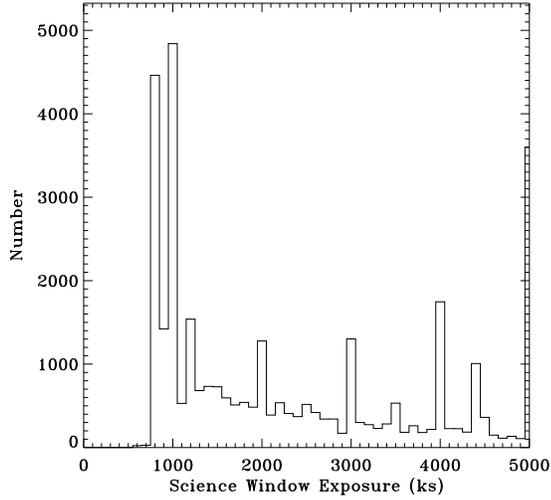}
\caption{\footnotesize {\it Top: }
The distribution of the exposure times for the UVW1 filter.
Other filters have not been used as much as the UVW1 filter,
so their values will be lower.
The solid line is the distribution of the mean exposure time
low-resolution mosaic (since the exposure time will vary over the mosaic)
and the dotted line is the distribution of exposure time for
the sum of the high resolution image centers.
{\it Bottom: }
The distribution of exposure times for individual science windows.
\label{fig:expo_stat}}
\end{figure}

\begin{figure}[h]
\epsscale{1.0}
\plotone{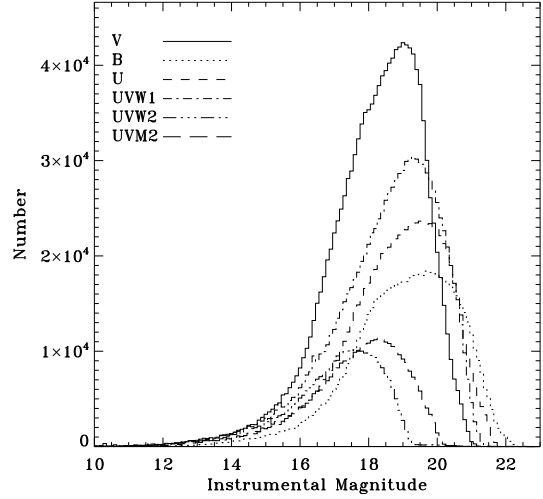}
\caption{\footnotesize The histogram of the numbers of sources at each magnitude
for each of the filters.
Only sources with $>3\sigma$ detections have been counted.
\label{fig:stat_mags}}
\end{figure}

Figure~\ref{fig:expo_stat} shows the distribution of total exposure time
per field and the exposure time per science window 
for the most commonly used filter (UVW1).
Most science windows have exposures of a kilosecond,
while the total exposure per field is significantly higher.
Thus, since the point source detection is done
in the individual science windows,
the OMCat could be made significantly deeper
were point source detection to be executed on mosaicked images.
As noted above,
since the bulk of individual science window exposures are $\sim1000$ seconds,
the detection depth of the catalogue is relatively uniform.
However, given the multiple coverage of sources
by individual science windows, the actual measurment accuracy
is significantly better than that obtained from a single 1000 second exposure.
Figure~\ref{fig:stat_mags} shows the distribution of magnitudes for each filter.

\begin{figure}[h!]
\epsscale{1.0}
\plotone{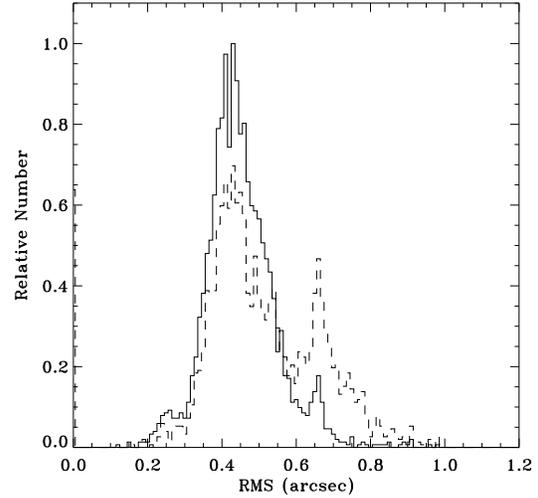}
\caption{\footnotesize The histogram of the resultant R.M.S. residuals
between the corrected source catalogue and the USNO-B1 catalogue
for those fields for which a coordinate correction was successful.
{\it Solid: } Relative number of fields after the post-pipeline correction.
{\it Dashed: } after the pipeline correction.
\label{fig:stat_resid}}
\end{figure}

{\it Coordinate Statistics:}
Figure~\ref{fig:stat_resid} shows the distribution of the residuals
between the corrected source catalogues and the USNO-B1 catalogue
for those fields for which a coordinate correction was successful.
The residuals before the post-pipeline correction 
are peaked around $0\farcs4$
with a secondary peak around $0\farcs7$.
After correction, 
the distribution is more symmetrically distributed around $0\farcs4$.
It should be noted that a portion of the residual for any given image
can be due to the offset of individual science windows within the mosaic.

\begin{figure}[h]
\epsscale{1.0}
\plotone{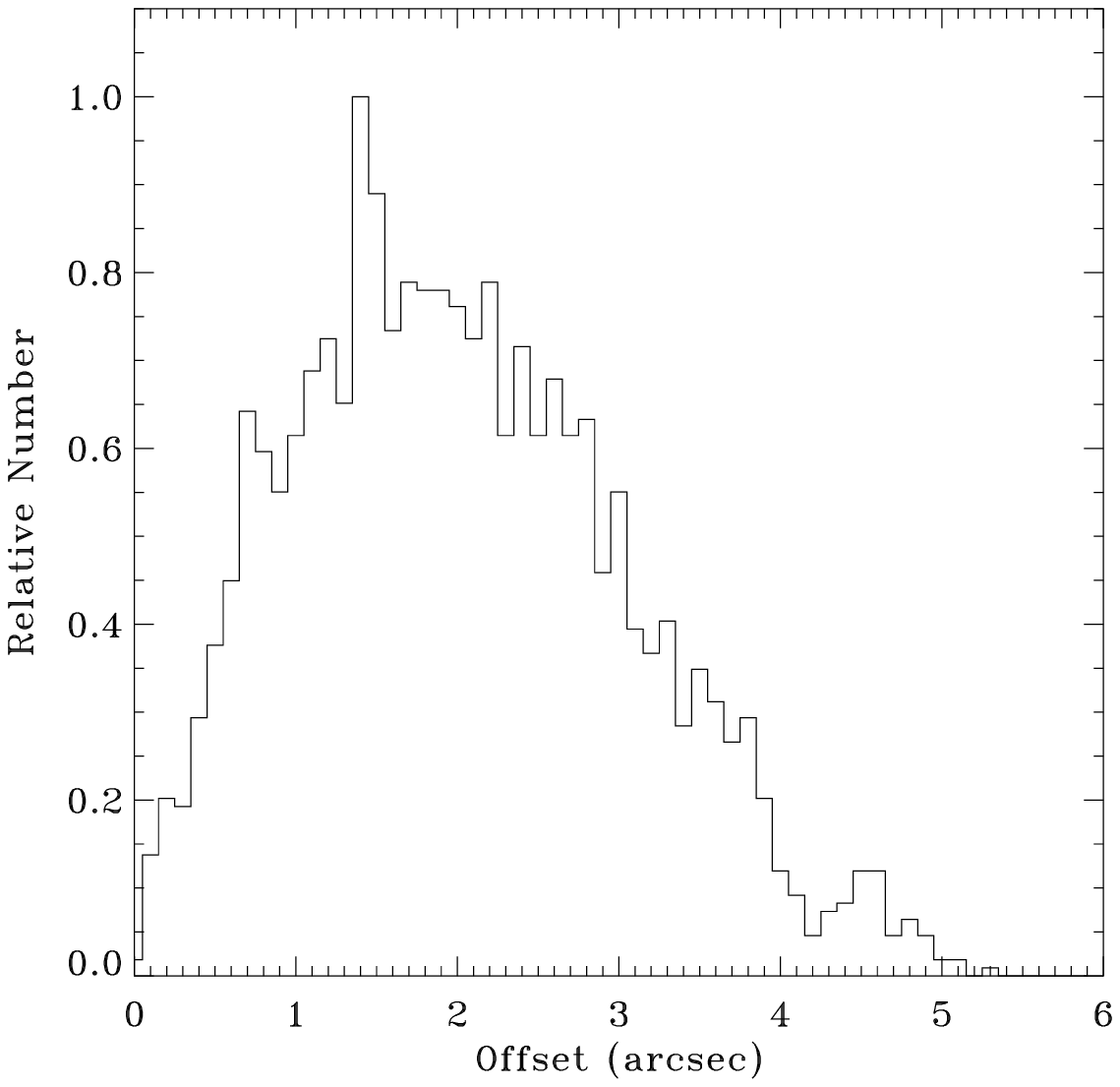}
\caption{\footnotesize The histogram of the magnitude of the shifts
introduced by the correction to USNO-B1 coordinates.
\label{fig:stat_offs}}
\end{figure}

The distribution of the calculated offsets between coordinates
before and after the post-pipeline correction by the USNO catalogue
is shown in Figure~\ref{fig:stat_offs}.
The distribution is a broad skewed Gaussian
peaking at $\sim1\farcs7$ 
with a significant tail extending to $\gtrsim5\arcsec$.
Note that this distribution does not reflect
the pointing ability of the telescope,
but rather the performance of the tracking corrections in the OM pipeline.

{\it Source Statistics:}
The OMCat contains roughly 3.7$\times10^6$ entries,
of which 82\% have detection significance 
(a maximum likelihood measure calculated within {\it omichain}) 
greater than three
in at least one band
and 72\% have a significance greater than three in at least one band.
Roughly 71\% of all entries have had successful coordinate corrections.
Only $\sim4\%$ of sources are classified as extended;
the rest are considered point-like.
Approximately 60\% of the sources 
are flagged for data quality in at least one band,
and approximately 74\% of the sources 
are flagged for confusion in at least one band.
However, if one looks only at individual measurements
for which the signal-to-noise ratio is $>3$,
49\% are flagged for data quality,
3\% are flagged for confusion,
and 4\% are flagged as extended.
Given the discussion in the previous section,
it is clear that the data quality flag in the current processing
is a very poor measure of the true data quality,
so these statistics, though they give one pause,
are not cause for alarm.

\begin{deluxetable}{lccccccc}
\tablecolumns{8}
\tabletypesize{\footnotesize}
\tablecaption{OM Color Statistics:
Number of 3$\sigma$ Detections\tablenotemark{a}
\label{tab:color_stats}}
\tablewidth{0pt}
\tablehead{
\colhead{Filter} &
\colhead{V} &
\colhead{B} &
\colhead{U} &
\colhead{UVW1} &
\colhead{UVM2} &
\colhead{UVW2} &
\colhead{WHITE} }
\startdata
V     & 1028785  & 233685 &  207608 &  165249 &   79448 &   56517 &    1867 \\
B     &  233685  & 548128 &  225180 &  133163 &   60190 &   66840 &    4600 \\
U     &  207608  & 225180 &  651438 &  247353 &   84993 &   62469 &    3546 \\
UVW1  &  165249  & 133163 &  247353 &  768349 &  144107 &   76204 &    2807 \\
UVM2  &   79448  &  60190 &   84993 &  144107 &  256142 &   79766 &     950 \\
UVW2  &   56517  &  66840 &   62469 &   76204 &   79766 &  191067 &     359 \\
WHITE &    1867  &   4600 &    3546 &    2807 &     950 &     359 &   11095 \\
\cutinhead{X-ray Selected Sources}
V    & 8536 & 2585 & 2732 & 2720 &  644 &  491 &   56 \\
B    & 2585 & 9694 & 4978 & 3964 &  803 &  685 &  403 \\
U    & 2732 & 4978 &15278 & 7016 & 1451 &  864 &  375 \\
UVW1 & 2720 & 3964 & 7016 &25477 & 2829 & 1852 &  334 \\
UVM2 &  644 &  803 & 1451 & 2829 & 6459 & 1467 &  143 \\
UVW2 &  491 &  685 &  864 & 1852 & 1467 & 7521 &   70 \\
WHITE&   56 &  403 &  375 &  334 &   70 &  143 &  812 \\
\enddata
\tablenotetext{a}{For colors, the requirement was that both
detections be greater than $3\sigma$,
not that the color be measured to a signal-to-noise of 3.}
\end{deluxetable}

Due to the interests of observers
and the changing default filter priorities,
the distribution of exposures among the various filters is uneven.
When measured in terms of the number of sources with $3\sigma$ detections
in each filter, the UVW1$-$U, B$-$V, and U$-$Bcolors have the best statistics.
Table~\ref{tab:color_stats}
shows the number of sources with $3\sigma$ detections
for each filter and filter combination.
As one might expect, the distribution for X-ray selected sources
(that is, X-ray point sources in the field with OMCat counterparts
rather than just sources observed because of their X-ray properties)
is somewhat different,
with UVW1$-$U, U$-$B, UVW1$-$B, and UVW2$-$UVW1 having the best statistics.

\begin{figure*}
\epsscale{1.0}
\centerline{\plotone{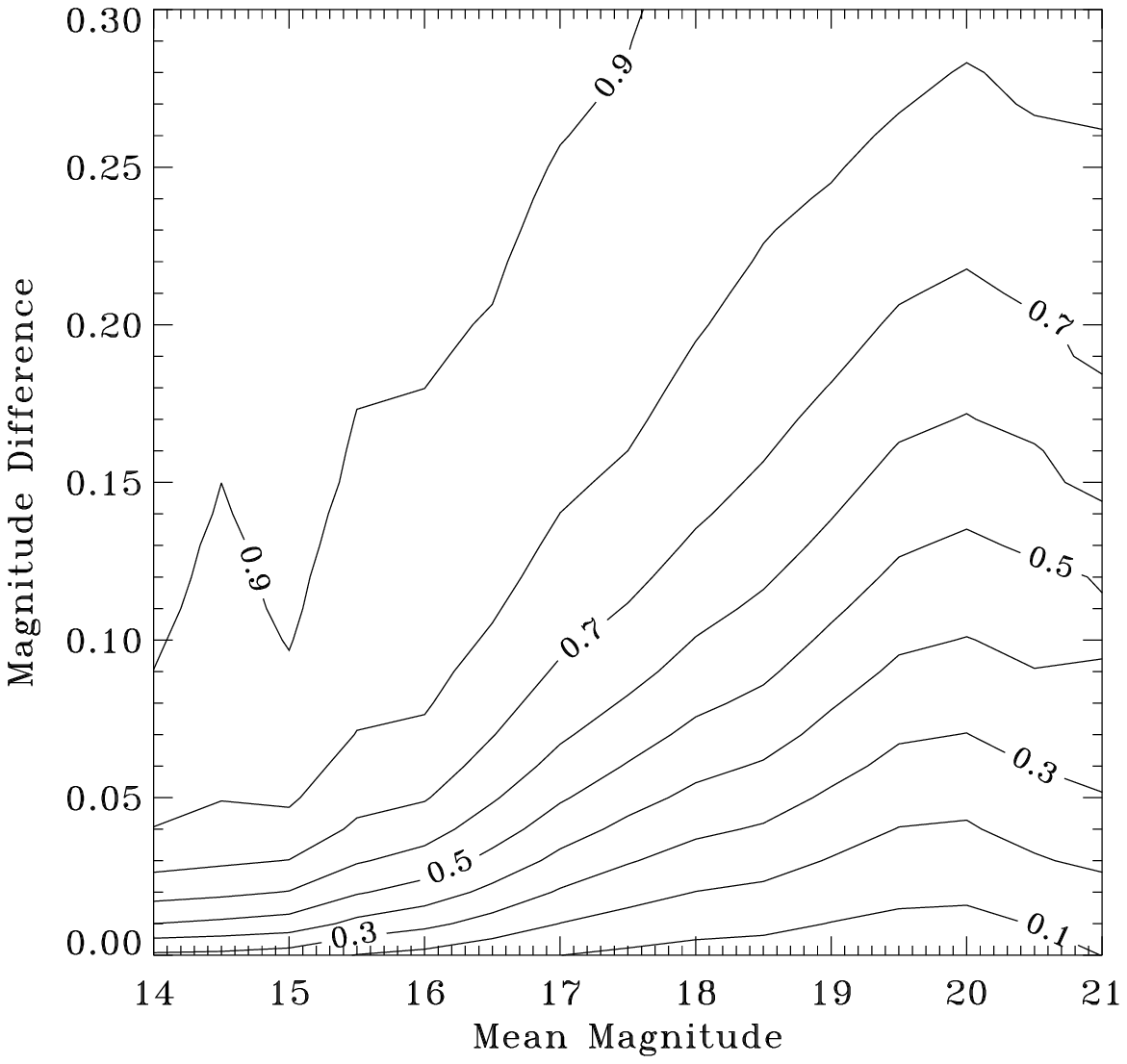}\plotone{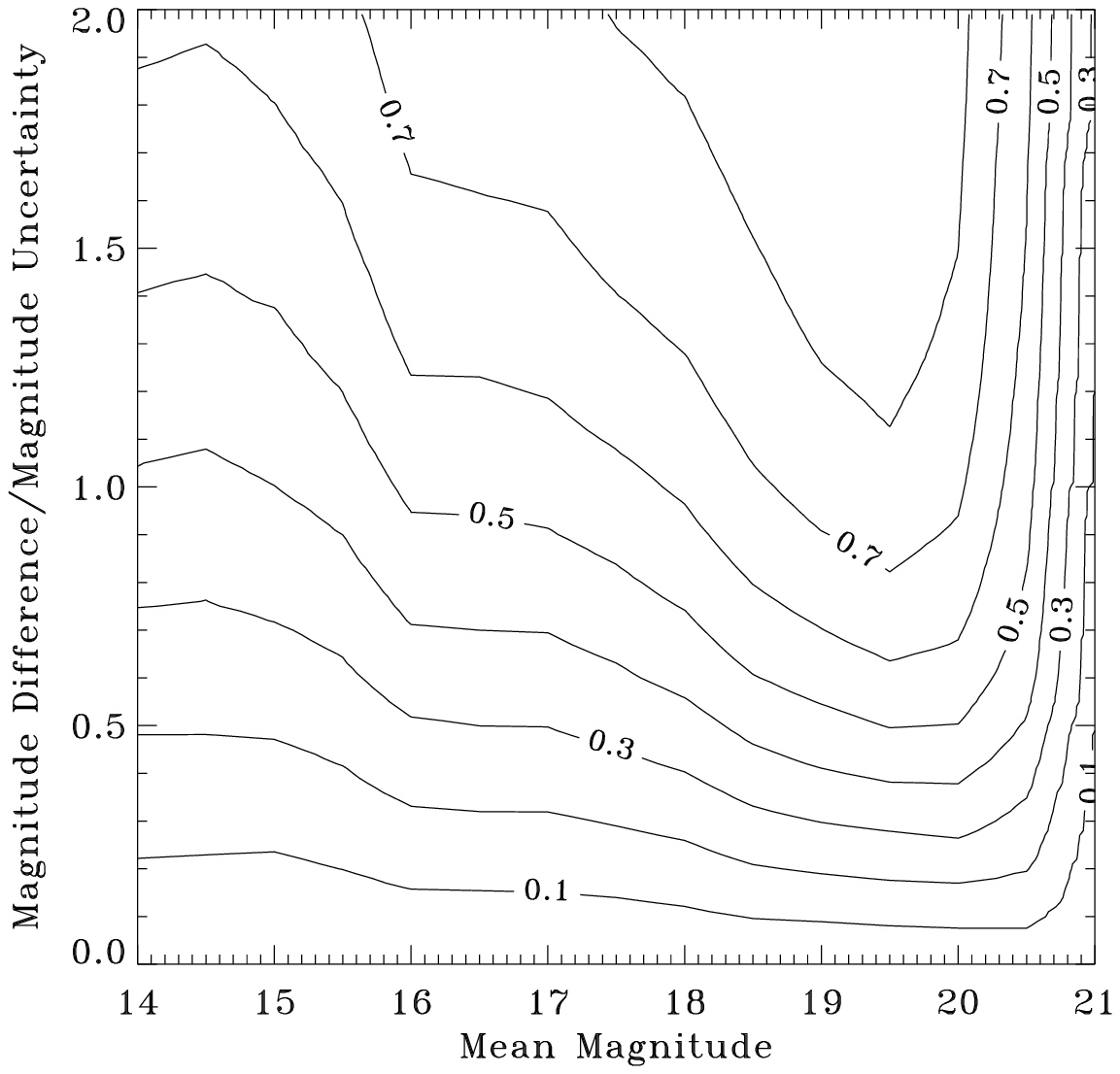}}
\caption{\footnotesize Two plots showing the photometric stability as a function of magnitude
as determined by pairs of measurements of the same object.
Some 199000 pairs of sources (not entirely unique source pairs)
were measured in the UVW1 filter.
Plots for the other filters show similar patterns.
{\bf Left:} The number of measured pairs having a given mean magnitude
and a given difference in magnitude shown as a contour diagram.
{\bf Right:} The distribution of magnitude difference
divided by its uncertainty as a function of mean magnitude.
The contours represent the fraction of pairs
at a given mean magnitude
with a difference/uncertainty ratio less than the value shown
on the vertical axis.
\label{fig:phot_stab}}
\end{figure*}

\begin{deluxetable}{lccccccc}
\tablecolumns{8}
\tabletypesize{\footnotesize}
\tablecaption{Photometric Reproducibility
\label{tab:phot_stab}}
\tablewidth{0pt}
\tablehead{
\colhead{Magnitude} &
\colhead{V} &
\colhead{B} &
\colhead{U} &
\colhead{UVW1} &
\colhead{UVM2} &
\colhead{UVW2} &
\colhead{White} }
\startdata
\cutinhead{Fraction of Pairs with $|M_1-M_2|<0.2$}
14 & 0.96 & 0.81 & 0.96 & 0.94 & 0.89 & 0.93 & 1.00 \\
15 & 0.96 & 0.95 & 0.94 & 0.93 & 0.93 & 0.90 & 0.87 \\
16 & 0.92 & 0.95 & 0.92 & 0.89 & 0.85 & 0.75 & 0.93 \\
17 & 0.91 & 0.91 & 0.91 & 0.86 & 0.79 & 0.58 & 0.95 \\
18 & 0.85 & 0.88 & 0.89 & 0.80 & 0.63 & 0.58 & 0.87 \\
19 & 0.74 & 0.84 & 0.82 & 0.73 & 0.60 & ...  & 0.88 \\
20 & 0.61 & 0.73 & 0.69 & 0.65 & ...  & ...  & ...  \\
21 & 0.54 & 0.66 & 0.66 & 0.72 & ...  & ...  & ...  \\
\cutinhead{Fraction of Pairs with $|M_1-M_2|<0.1$}
14 & 0.94 & 0.79 & 0.92 & 0.91 & 0.82 & 0.88 & 1.00 \\
15 & 0.92 & 0.89 & 0.89 & 0.88 & 0.84 & 0.75 & 0.85 \\
16 & 0.87 & 0.88 & 0.86 & 0.82 & 0.66 & 0.53 & 0.91 \\
17 & 0.78 & 0.83 & 0.82 & 0.70 & 0.55 & 0.36 & 0.87 \\
18 & 0.63 & 0.75 & 0.69 & 0.57 & 0.37 & 0.33 & 0.84 \\
19 & 0.45 & 0.61 & 0.55 & 0.47 & 0.32 & ...  & 0.79 \\
20 & 0.34 & 0.44 & 0.41 & 0.37 & ...  & ...  & ...  \\
21 & 0.38 & 0.38 & 0.39 & 0.39 & ...  & ...  & ...  \\
\cutinhead{Fraction of Pairs with $|M_1-M_2|<0.05$}
14 & 0.80 & 0.73 & 0.82 & 0.81 & 0.61 & 0.70 & 0.00 \\
15 & 0.81 & 0.74 & 0.77 & 0.74 & 0.64 & 0.47 & 0.79 \\
16 & 0.70 & 0.72 & 0.71 & 0.65 & 0.40 & 0.30 & 0.88 \\
17 & 0.53 & 0.65 & 0.62 & 0.46 & 0.31 & 0.19 & 0.80 \\
18 & 0.37 & 0.53 & 0.42 & 0.33 & 0.20 & 0.17 & 0.78 \\
19 & 0.24 & 0.35 & 0.31 & 0.25 & 0.16 & ...  & 0.58 \\
20 & 0.18 & 0.23 & 0.22 & 0.19 & ...  & ...  & ...  \\
21 & 0.12 & 0.20 & 0.21 & 0.27 & ...  & ...  & ...  \\
\enddata
\end{deluxetable}

{\it Photometric Statistics:}
From the OMCat one can empirically determine 
the repeatability of flux measurements as a function of source flux as,
for example, by \citet{antokhin} to study the sources of systematic errors.
If the same source is measured twice for a given filter
and produces measurements of $M_1$ and $M_2$,
then Figure~\ref{fig:phot_stab} (left) shows the distribution of 
the magnitude difference $|M_1-M_2|$
as a function of the mean magnitude $(M_1+M_2)/2$.
One can see that the bulk of the 199000 pairs measured have 
$|M_1-M_2|<0.5$ magnitudes, though there is a long tail to much higher values,
in part due to intrinsic variability and 
(especially in more crowded fields) source matching errors.
Figure~\ref{fig:phot_stab} (right) shows the distribution of
$|M_1-M_2|/\sqrt{\sigma_1^2 + \sigma_2^2}$
as a function of $(M_1+M_2)/2$.
If the difference in magnitudes were strictly statistical,
the contours in this diagram would be strictly horizontal
with the 68\% contour at the y-axis value of unity.
Table~\ref{tab:phot_stab} shows the fraction of pairs
with $|M_1-M_2|<$ 0.2, 0.1, and 0.05 magnitudes as a function of magnitude
for each of the filters.
As can be seen in Figure~\ref{fig:phot_stab},
excluding the very brightest and very faintest sources,
the $|M_1-M_2|$ is roughly what one would expect from the quoted uncertainties.

{\it Comparison to the GALEX AIS:}
Perhaps the most comparable UV survey is the {\it GALEX}
AIS, an ``all sky survey'' which, when completed,
will cover $\sim50$\% of the sky in its NUV band
to a limiting AB magnitude of 20.8 for a $5\sigma$ detection
\citep{galex_martin,galex_morrissey}.
The closest OM equivalent to the {\it GALEX} NUV filter is the UVM2 filter.
The magnitude distribution of {\it GALEX} NUV sources with signal-to-noise ratios $>5$
is sharply peaked at 20.8 and drops to 10\% of the peak value by a magnitude of 21.4.
The magnitude distribution of OM sources in the UVM2 filter
with signal-to-noise ratios $>5$ is more broadly peaked
(due to the range of exposure times),
with a peak at an instrumental magnitude of $\sim16.75$
and declines to 10\% of the peak value by an instrumental magnitude of $\sim18.25$.
The equivalent AB magnitudes are $\sim18.4$ and $\sim19.9$ respectively.
Thus, the current OMCat has a limiting magnitude in UVM2 
between 1.5 and 2.5 magnitudes brighter
than the limiting magnitude in the {\it GALEX} NUV filter.
(However, it should be noted that because of the smaller FOV
and different brightness limits, the OMCat has good coverage of the Galactic plane,
which {\it GALEX} is not able to observe.)

{\it Comparison to the SDSS:}
The SDSS u filter, like the OMCat U filter, is similar to the Johnson U.
The magnitude distribution of the SDSS u band sources 
with signal-to-noise ratios $>5$
is sharply peaked at 21.5 
and drops to 10\% of the peak value by a magnitude of 22.4.
The  magnitude distribution of OM sources in the U filter
with signal-to-noise ratios $>5$ is more broadly peaked
with a peak at an instrumental magnitude of $\sim18.75$
and declines to 10\% of the peak value 
by an instrumental magnitude of $\sim20.5$.
The equivalent AB magnitudes are 19.7 and 21.4.
Thus the current OMCat has a limiting magnitude in U
about 2 magnitudes brighter than the SDSS.

\section{Source Identification in the OM Color-color Planes}

\begin{figure*}
\epsscale{1.0}
\plotone{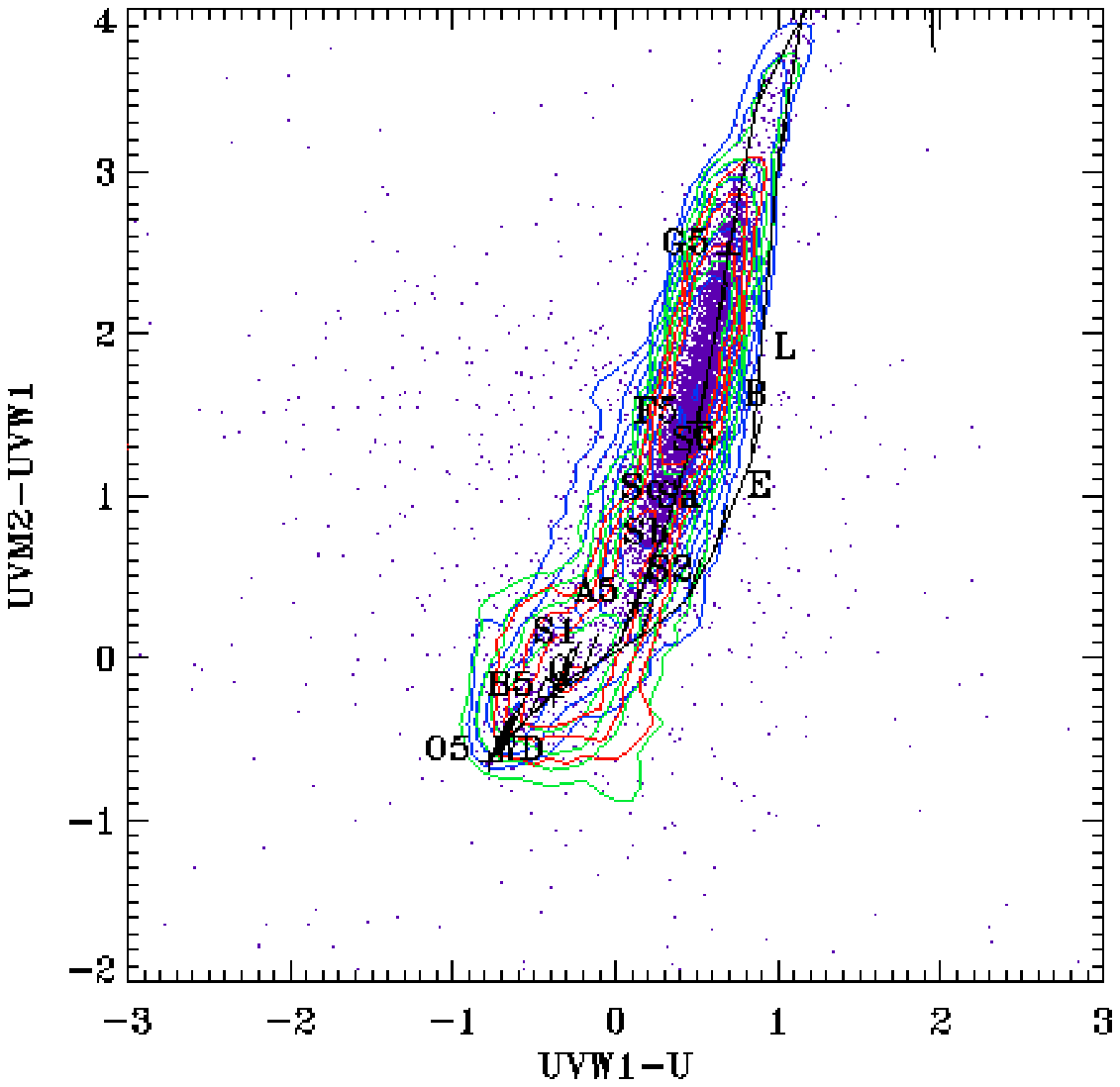}
\epsscale{1.0}
\plotone{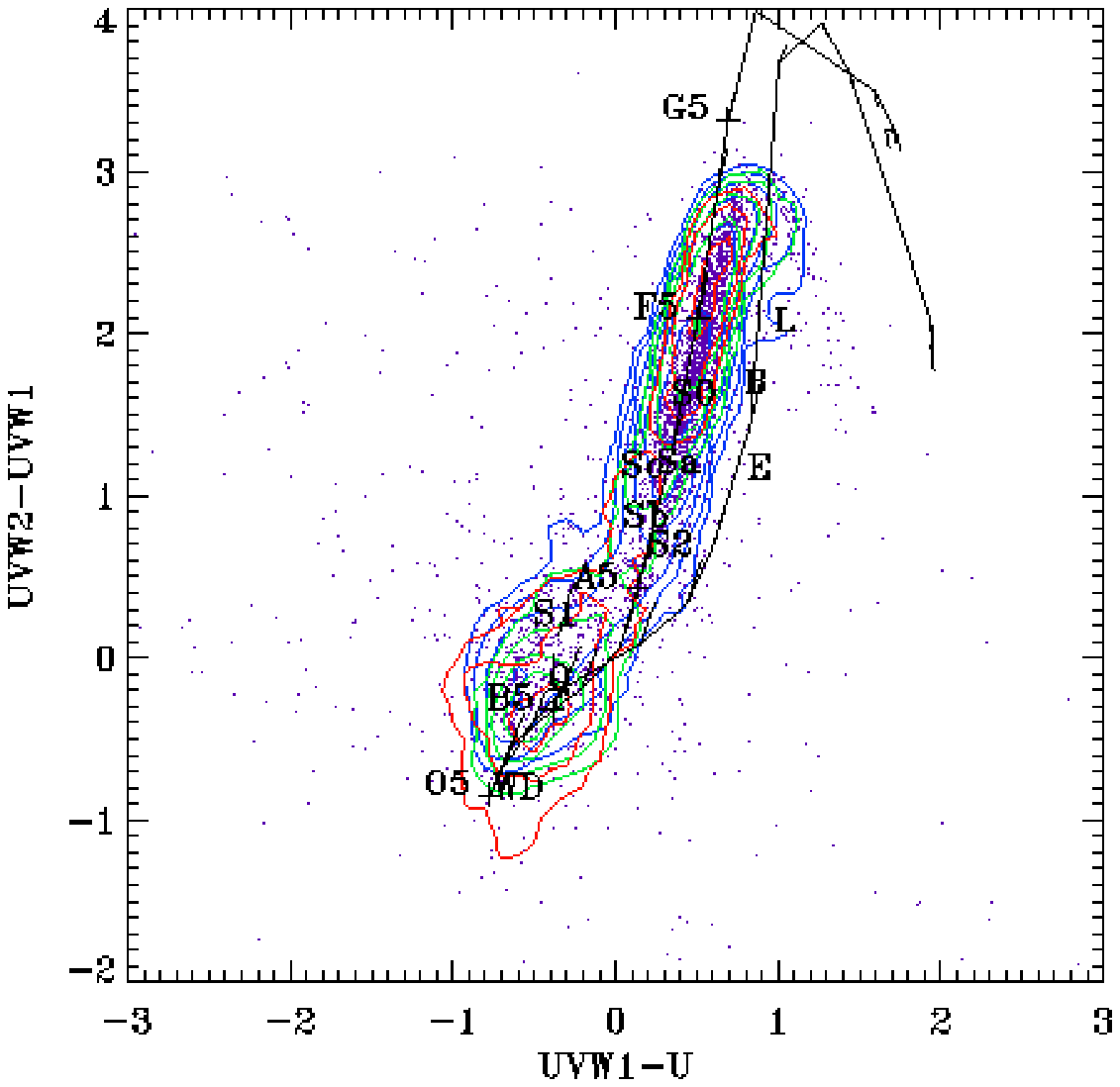}
\epsscale{1.0}
\plotone{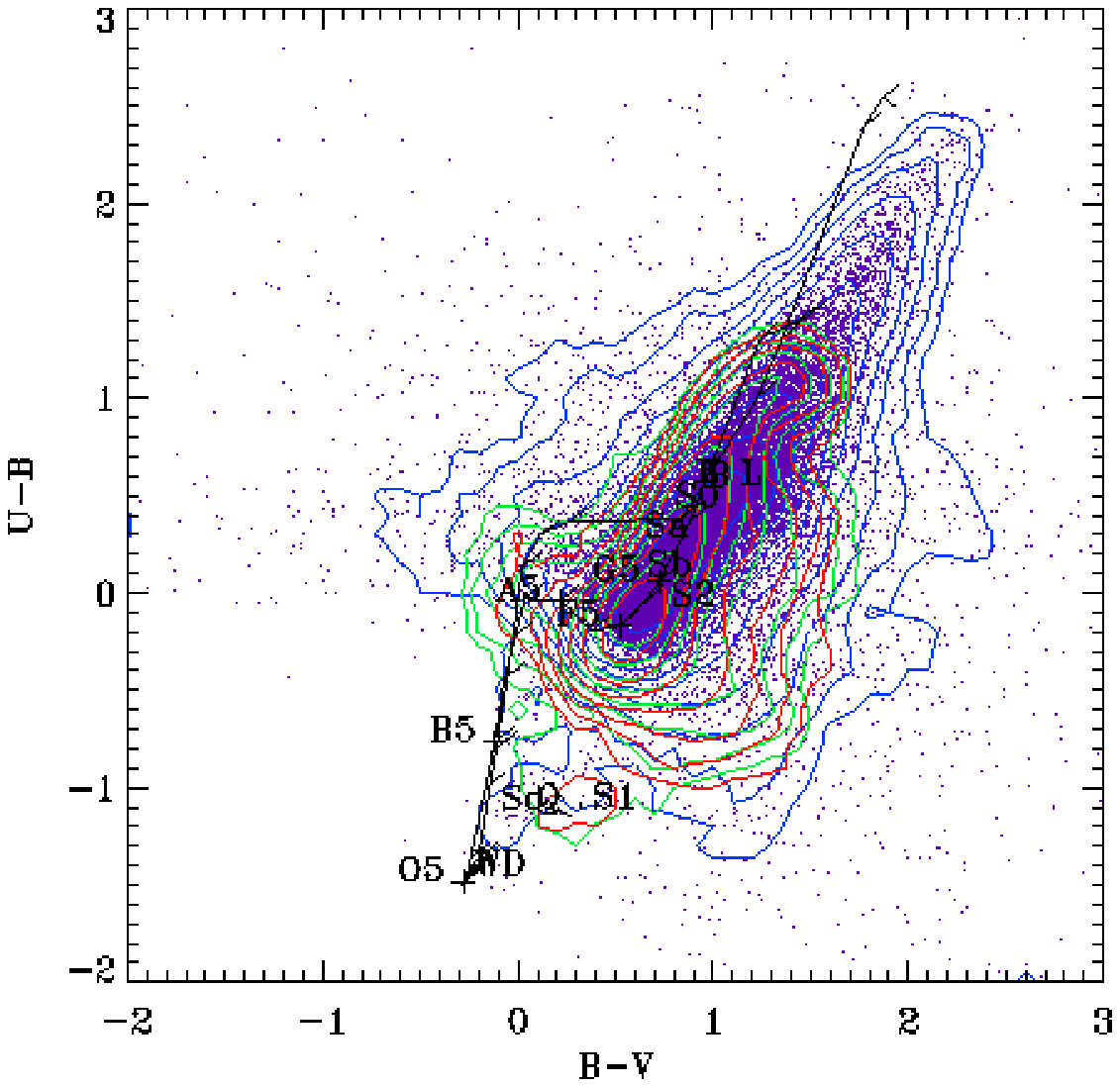}
\caption{\footnotesize Color-color diagrams for the OM filters.
Only sources with significances greater than 3
and $\sigma_{mag}<0.15$ in each filter have been plotted.
The thick solid line upon which the spectral types are indicated
is the track of solar-metallicity dwarfs
plotted with representative spectral types.
The thin lines are reddening vectors for each spectral type
assuming E(B-V)=0.1 and the \citet{fitz1999} reddening curve.
No reddening corrections have been applied to the data.
The other track (without spectral type indicators)
is for solar-metallicity giants.
Also marked are the locations of white dwarfs (WD),
QSOs (Q), Seyfert 1 and 2 (S1 and S2), LINERs (L)
typical galaxies (S0, Sa, Sb, Sc, and E)
as well as spiral bulges (B).
The {\it blue} contours are from fields with $|b|<30\arcdeg$,
{\it green} contours from fields with $30\arcdeg<|b|<60\arcdeg$,
and {\it red} contours from fields with $60\arcdeg<|b|$.
The purple points are the sources from which the blue contours were calculated.
\label{fig:color_color}}
\end{figure*}

The OM unique filters are UVW1, UVM2, and UVW2, 
in order of decreasing throughput (see Figure~\ref{fig:filts}).
In order to explore the abilities of the OM filters,
we created UVW1$-$U {\it vs.} UVM2$-$UVW1 and 
UVW1$-$U {\it vs.} UVW2$-$UVW1 color-color diagrams 
of the sources in the OMCat (Figure~\ref{fig:color_color}).
On those diagrams we have plotted the expected locus for main-sequence stars
(solar metallicity stars from the 1993 Kurucz 
atlas\footnote{http://www.stsci.edu/hst/observatory/cdbs/k93models.html}),
as well as points representative of galaxies
\citep[taken from the Kinney \& Calzetti atlas at 
STScI\footnote{http://www.stsci.edu/hst/observatory/cdbs/cdbs\_kc96.html}
see][]{kea1996}
and AGN \citep[taken from STScI AGN atlas collection of 
spectra\footnote{http://www.stsci.edu/hst/observatory/cdbs/cdbs\_agn.html}
see][]{fea1991}.
The conversion from spectra to photometric colors was made using
the OM spectral response matrices available from the \xmm\ 
SOC\footnote{ftp://xmm.esac.esa.int/pub/ccf/constituents/extras/responses/OM}.

It is immediately apparent that the OM colors
are a good match to those expected from the Kurucz atlas,
except for the stars later than about G5.
This problem appears most clearly in the UVW1$-$U {\it vs.} UVW2$-$UVW1 diagram,
but appears as well in the B$-$V {\it vs.} U-B diagram.
The problem region is off the bottom of the UVW1$-$U {\it vs.} UVM2$-$UVW1 diagram.
The true nature of the problem is not yet clear.
For much of the stellar tracks the reddening vectors 
are roughly parallel to the stellar track.

\begin{figure*}
\epsscale{1.0}
\plotone{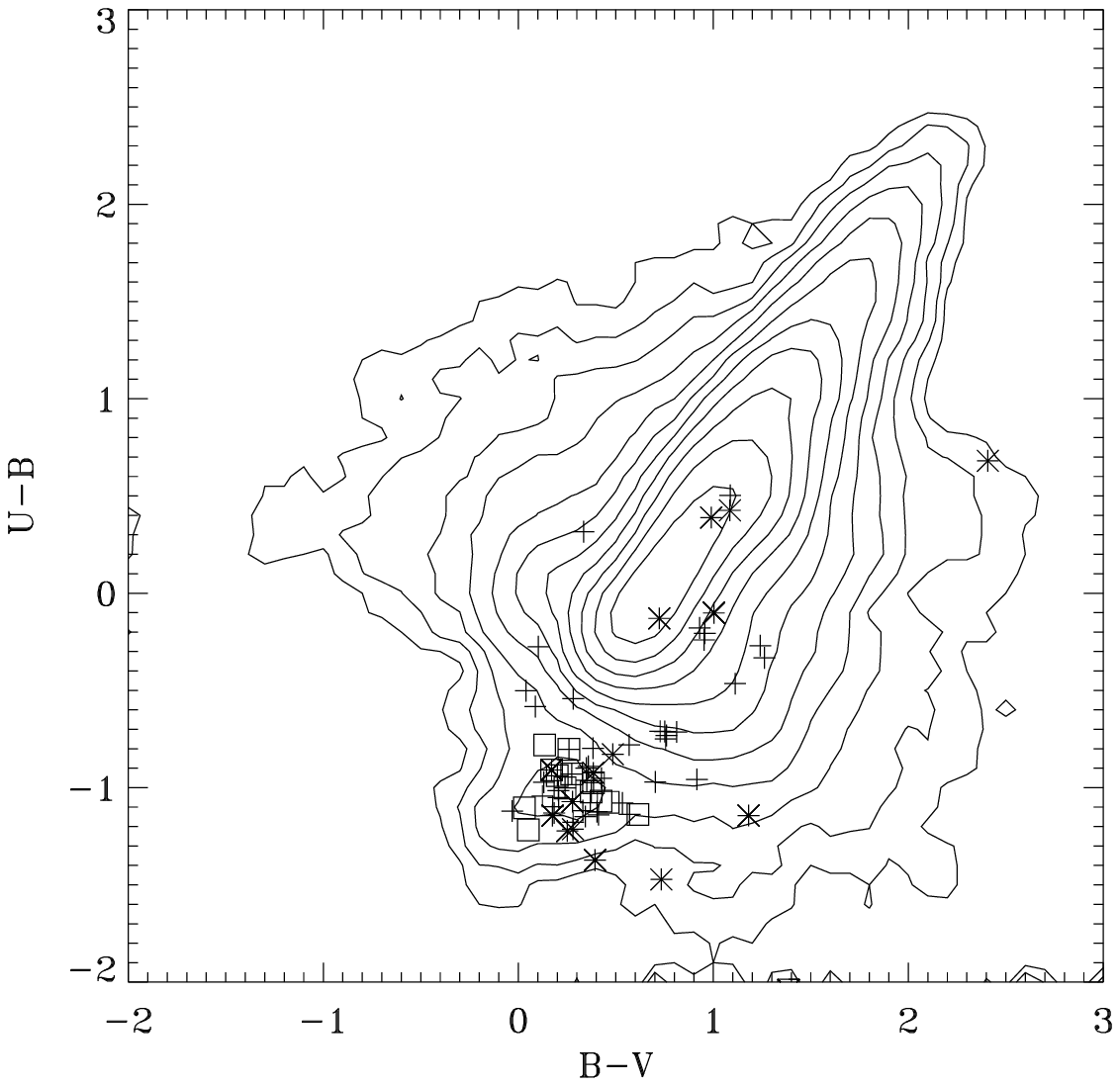}
\epsscale{1.0}
\plotone{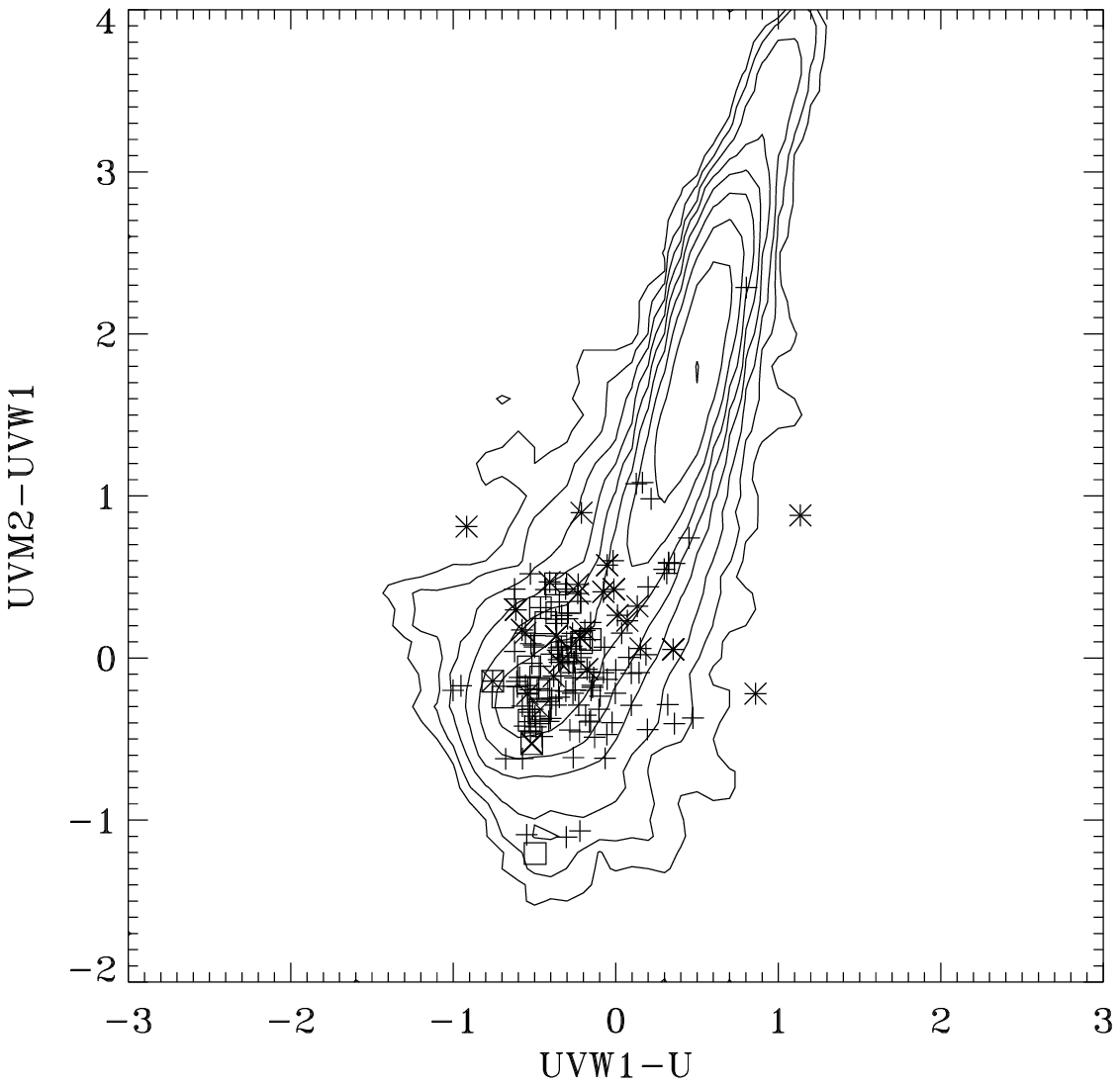}
\caption{\footnotesize The contours of the source density of all OMCat sources
with detection significance $>3$ and $\sigma_{mag}<0.3$.
The contours are logarithmically spaced.
The location of QSO from the 12$^{th}$ Veron catalogue ({\it $+$}),
the QSO from the SDSS catalogue ($\Box$),
and the a mixed sample of QSO and galaxies that is the CfA redshift survey
($\times$).
There is a great deal of overlap between the Veron catalogue
and the CfA redshift survey.
\label{fig:qso_color}}
\end{figure*}

There is a strong overlap between the early dwarf stars (down to A5) and AGN,
as well as somewhat later dwarf stars (A5 to F5) and typical galaxies.
There is no way to distinguish early stars from AGN using the UV colors alone,
and the UV colors of QSO do not change much with redshift.
The AGN seem to be more offset from the early-type stars
in the optical color-color diagram.
Given the extinction in the UV,
it would be unreasonable to expect to find AGN 
in regions with early type stars,
though the converse is not necessarily true,
and indeed, finding high latitude O and B stars would be interesting.
Division of sources into low ($|b|<30\arcdeg$) 
and high ($|b|>60\arcdeg$) Galactic latitude samples
shows that in the UV color-color diagram
the early type star region is populated somewhat more strongly 
at higher Galactic latitudes
while the later type star region is populated 
somewhat more strongly at lower Galactic latitudes.
We have cross-correlated the OMCat with 
the 12$^{th}$ Veron QSO catalogue \citep{veroncat}
and a SDSS AGN catalogue \citep{kea2003}
and found that the objects in common lie 
in the expected locations in the color-color diagram 
(Figure~\ref{fig:qso_color}).

\begin{figure*}
\epsscale{1.0}
\plotone{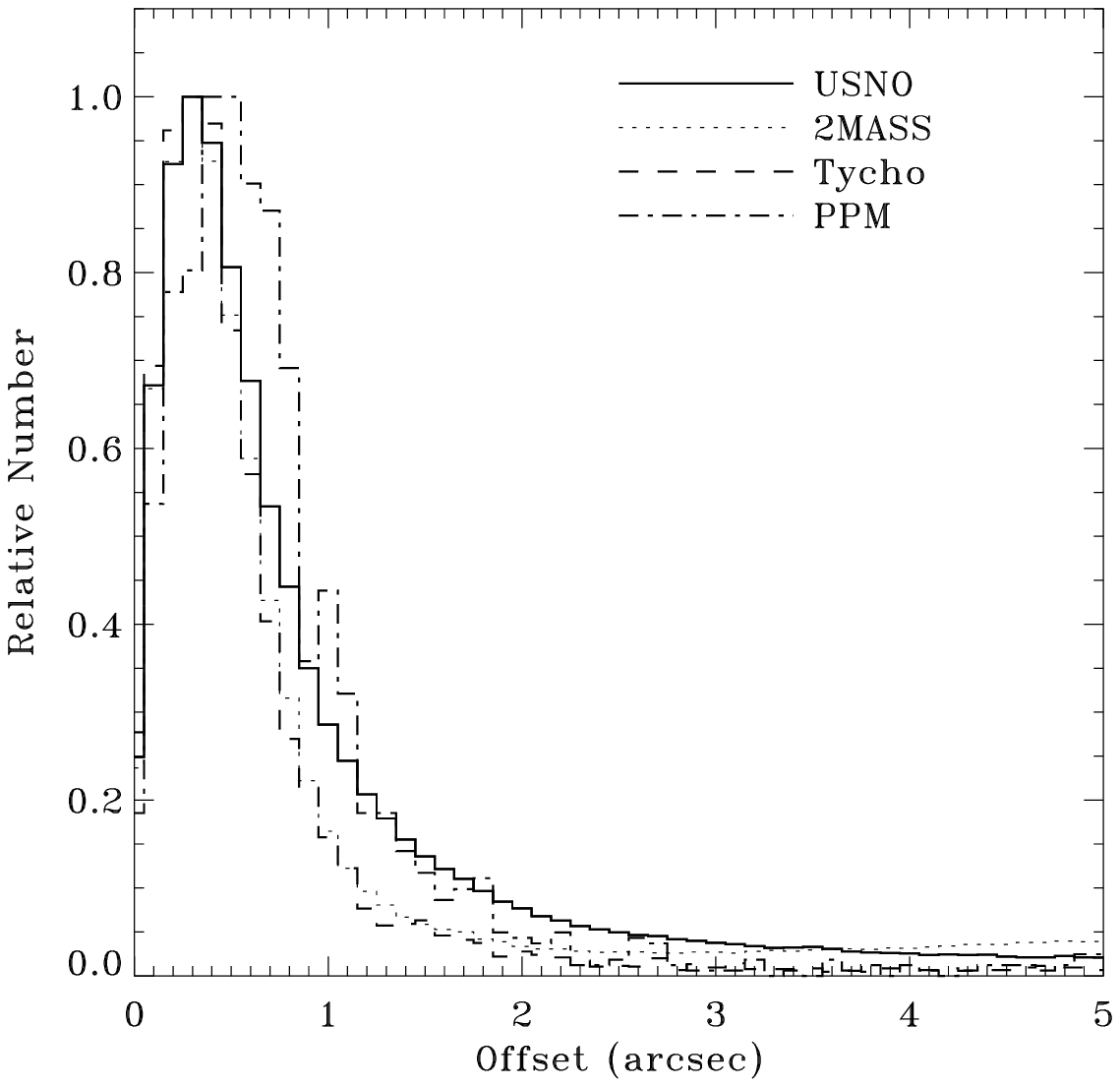}
\epsscale{1.0}
\plotone{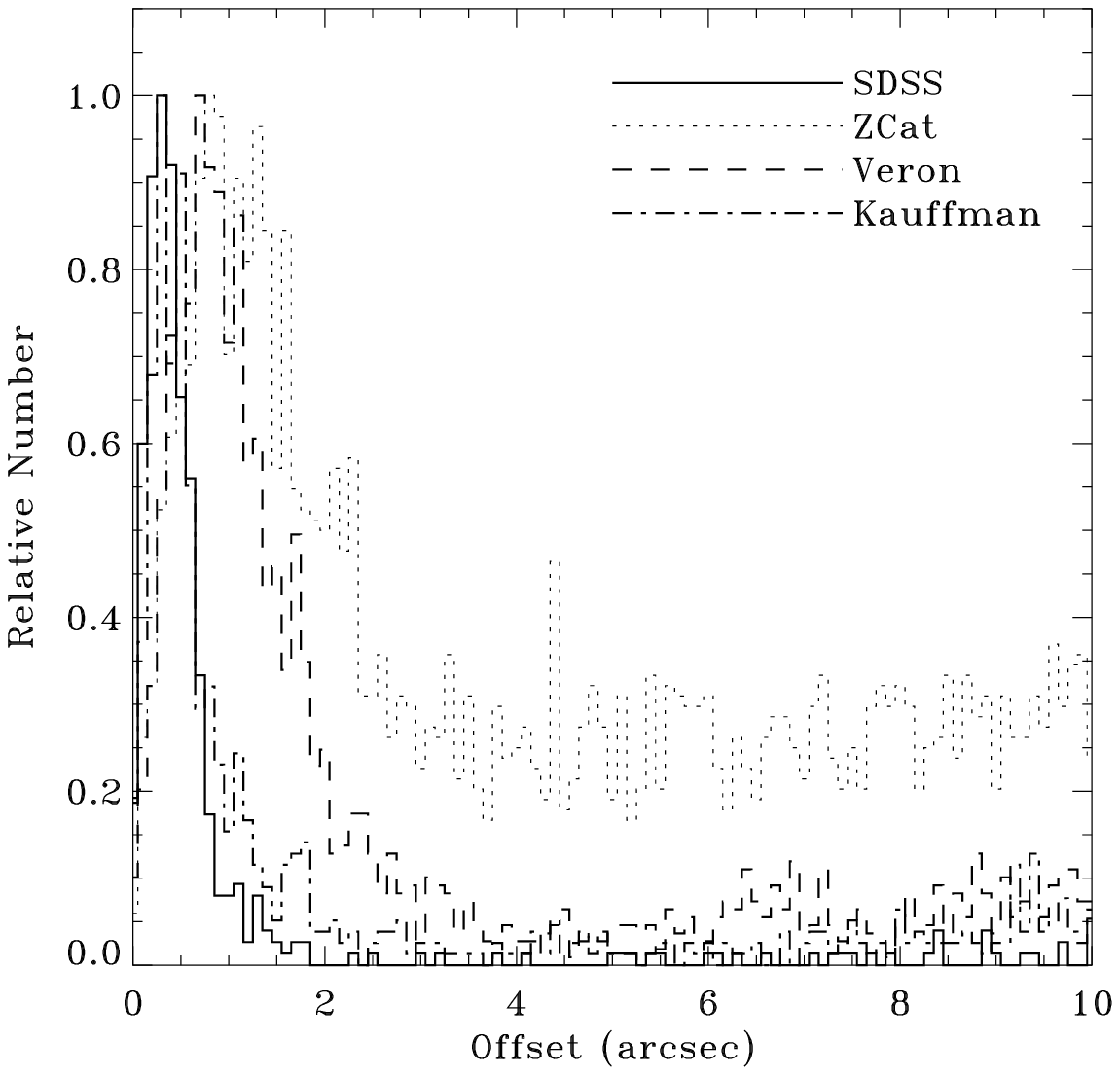}
\epsscale{1.0}
\plotone{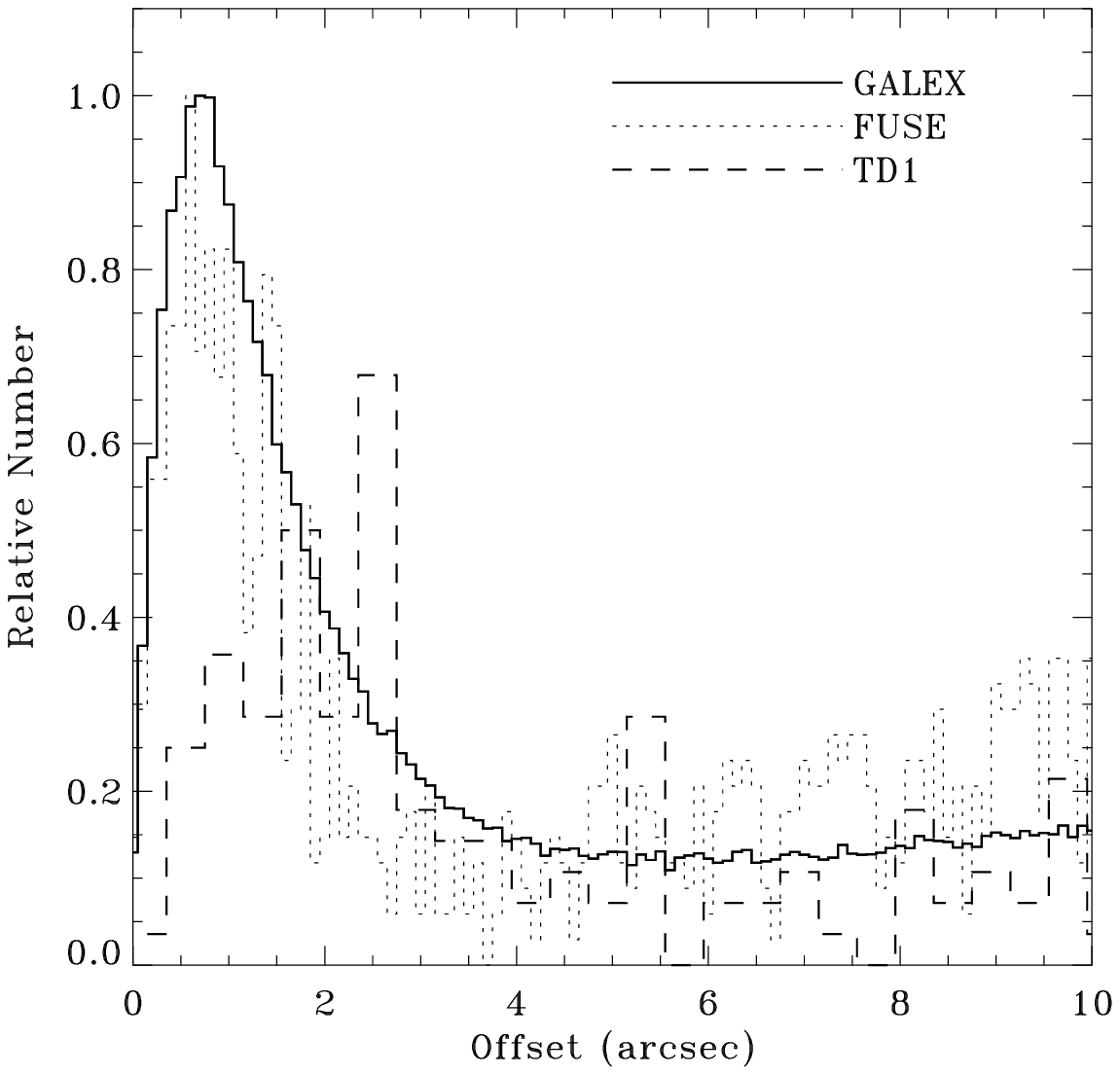}
\caption{\footnotesize Each diagram is the distribution of the distance
between sources in the OMCat
and the nearest source in some other catalogue.
{\bf Top Left: }Catalogues dominated by stars and galaxies.
{\bf Top Right: }Catalogues dominated by QSO and AGN.
{\bf Bottom: }Catalogues of UV sources.
Note that since {\it FUSE} does not measure source positions
(the positions were supplied by the user,
typically from catalogues such as the Guide Star Catalogue)
the width of the distribution for {\it FUSE} does not reflect
the intrinsic accuracy of the {\it FUSE} pointing.
The width of the distribution does still indicate
the radius at which spurious matches become important.
\label{fig:cross}}
\end{figure*}

As can be seen in Figure~\ref{fig:cross},
the cross-correlation of the CfA redshift survey 
\citep[the 1995 version as held by HEASARC,][]{zcat1992} with the OMCat
suggests that the sources falling within the matching radius
will be strongly contaminated by false-matches
(at the $\sim20$\% level),
meaning that many OMCat stellar sources will be falsely matched 
to the CfA survey.
This high false-matching rate is presumably due to the fact that either
the CfA redshift survey has a very high density of objects
that are much fainter than those detected in the OMCat,
or that the CfA redshift survey, 
being primarily a catalogue of extended objects,
have positions that are inherently less precise
than those in catalogues of true point sources.
However, the matching sources from the CfA redshift survey
fall at the expected location in the color-color diagrams,
though their scatter may be a bit larger than the other surveys.

\section{Uses of the OMCat}

\begin{deluxetable}{rrrcccc}
\tablecolumns{7}
\tabletypesize{\footnotesize}
\tablecaption{Cross-Correlations
\label{tab:cross}}
\tablewidth{0pt}
\tablehead{
\colhead{Catalogue} &
\colhead{Entries} &
\colhead{Matches} &
\colhead{Coordinate} &
\colhead{Match} &
\colhead{Match} &
\colhead{Match} \\
\colhead{Name} &
\colhead{} &
\colhead{} &
\colhead{Uncertainty\tablenotemark{a}} &
\colhead{Peak} &
\colhead{Width} &
\colhead{Radius} }
\startdata
USNO B-1      & 1045913669 & 455481 & $0\farcs2$                  & 0.303 & 0.199 & 0.702 \\
2Mass         &  470992970 & 265991 & $0\farcs06$                  & 0.255 & 0.162 & 0.580 \\
Palomar-Green &       1878 & \ldots & $8\arcsec$                  & \ldots & \ldots & \ldots \\
Tycho-2       &    2539913 &   5322 & $0.06$                      & 0.271 & 0.189 & 0.650 \\
PPM           &     468861 &   1177 & $0\farcs3$                  & 0.469 & 0.309 & 1.087 \\
SAO           &     258944 &    578 & $\sim2\arcsec$              & 0.920 & 0.478 & 1.876 \\
CfA Z Cat     &      58738 &   1366 & $1\arcsec$\tablenotemark{b} & 1.470 & 0.574 & 2.619 \\
Veron         &     108080 &   1099 & $1\arcsec$\tablenotemark{b} & 0.706 & 0.341 & 1.388 \\
SDSS NBC QSO  &     100563 &    395 & $0.36$mas\tablenotemark{b}  & 0.323 & 0.210 & 0.742 \\
GALEX         &  110236958 &  42528 & $1\arcsec$                  & 0.647 & 0.349 & 1.346 \\
FUSE          &       4037 &    376 & $1\farcs5$\tablenotemark{c} & 1.225 & 0.442 & 2.109\tablenotemark{c} \\
TD1           &      31215 &     82 & $0\farcs2$\tablenotemark{b} & 2.036 & 1.089 & 4.215 \\

\enddata
\tablenotetext{a}{From catalogue documentation accessed through the HEASARC.}
\tablenotetext{b}{Precision of coordinate as the uncertainty was not quoted.}
\tablenotetext{c}{Since the FUSE catalogue is of observed targets,
the coordinates are presumably those provided by the observers.
The smallest aperture in use has a width of $1\farcs5$
while the medium aperture has a width of $4\farcs0$,
suggesting that the coordinates are at least this good
in order to make a successful observation.}
\end{deluxetable}

The uses of the OMCat detailed here involve 
the cross-correlation of the OMCat with other catalogues.
All the catalogues were extracted from the HEASARC Catalogue Resources 
using the HEASARC Browse tool\footnote{
The USNO-B1.0 and 2MASS catalogues are maintained by the VizieR service 
of the Centre de Donn\'{e}es Astronomiques de Strasbourg
and were accessed through the HEASARC Browse interface.}.
The cross-correlation for a given catalogue was done by determining
the closest entry to each OMCat source.
We then plotted the distribution of the distances
between the OMCat source and the closest catalogue source.
Cross-correlation of the OMCat with catalogues with a high density of sources
typically produced a distribution of distances 
with a strong peak at $\sim0.2-0.3$ arcseconds and a FWHM of $\sim$0.3 arcseconds,
and a tail due to serendipitous matches;
the higher the density of catalogue sources,
the higher the serendipitous match rate at large distances.
An OMCat source was generally considered to be matched
if the distance between it and the catalogue source
was less than the peak of the distribution  $+2\sigma$.
The catalogues and match criteria used are shown in Table~\ref{tab:cross}
while the distributions are shown in Figure~\ref{fig:cross}.

\subsection{USNO Counterparts}
Most of the sources in the OMCat have USNO counterparts.
This suggests that the OMCat is shallower in the UV
than the USNO is in the B and R bands.
The OMCat is therefore not dominated by sources
previously undetected at other wavelengths.

\subsection{FUSE Counterparts}
Of interest to studies of the halo of the Galaxy 
are UV-bright stars and AGN that can be used as background sources
for measuring the column density, velocity, and metallicity of halo gas.
Stars are of particular use in determining the distance 
to high and intermediate velocity gas \citep[e.g.,][]{dak1993},
while AGN provide measures of the column density through the entire halo.
Targets are usually found by combing catalogues of sources
for objects with optical colors suggesting high UV fluxes.
Since the fields in the OMCat were (usually) chosen for their X-ray sources, 
rather than the UV sources in the same field,
comparing the UV sources in the OMCat with the types of catalogues
that have been used in the past to find UV-bright objects
provides an indication of how many UV-bright sources may have been missed.

We have cross-correlated the OMCat with several UV catalogues;
the {\it GALEX} catalogue, 
the Far Ultraviolet Explorer ({\it FUSE}) observation log
(not technically a {\it source} catalogue),
and the TD1 catalogue;
the offset distributions are shown in Figure~\ref{fig:cross}.
We also cross-correlated with the 
Far-UV Space Telescope ({\it Faust}) Far-UV Point Source Catalogue,
but did not find any matches, 
likely due to the small number of sources in that catalogue.
There are a number of other EUV catalogues 
from {\it EUVE} and the {\it ROSAT} WFC,
but the position uncertainties are $\gtrsim1\arcmin$,
so they are not useful for this study.

At the time that the first OMCat was constructed,
{\it FUSE} was still operational and could have been used
for follow-up spectroscopic studies.
Even with the demise of {\it FUSE},
using the {\it FUSE} criteria for finding UV sources of interest
to be studied with UV spectroscopy
has application for future missions/instruments 
such as the HST Cosmic Origins Spectrograph.
If searching the OMCat for {\it FUSE} observable sources 
produces a trivial result,
then the OMCat is not likely to be of interest for future missions;
a bountiful return would suggest that the OMCat
will be useful.

\begin{figure*}
\epsscale{1.0}
\centerline{\plotone{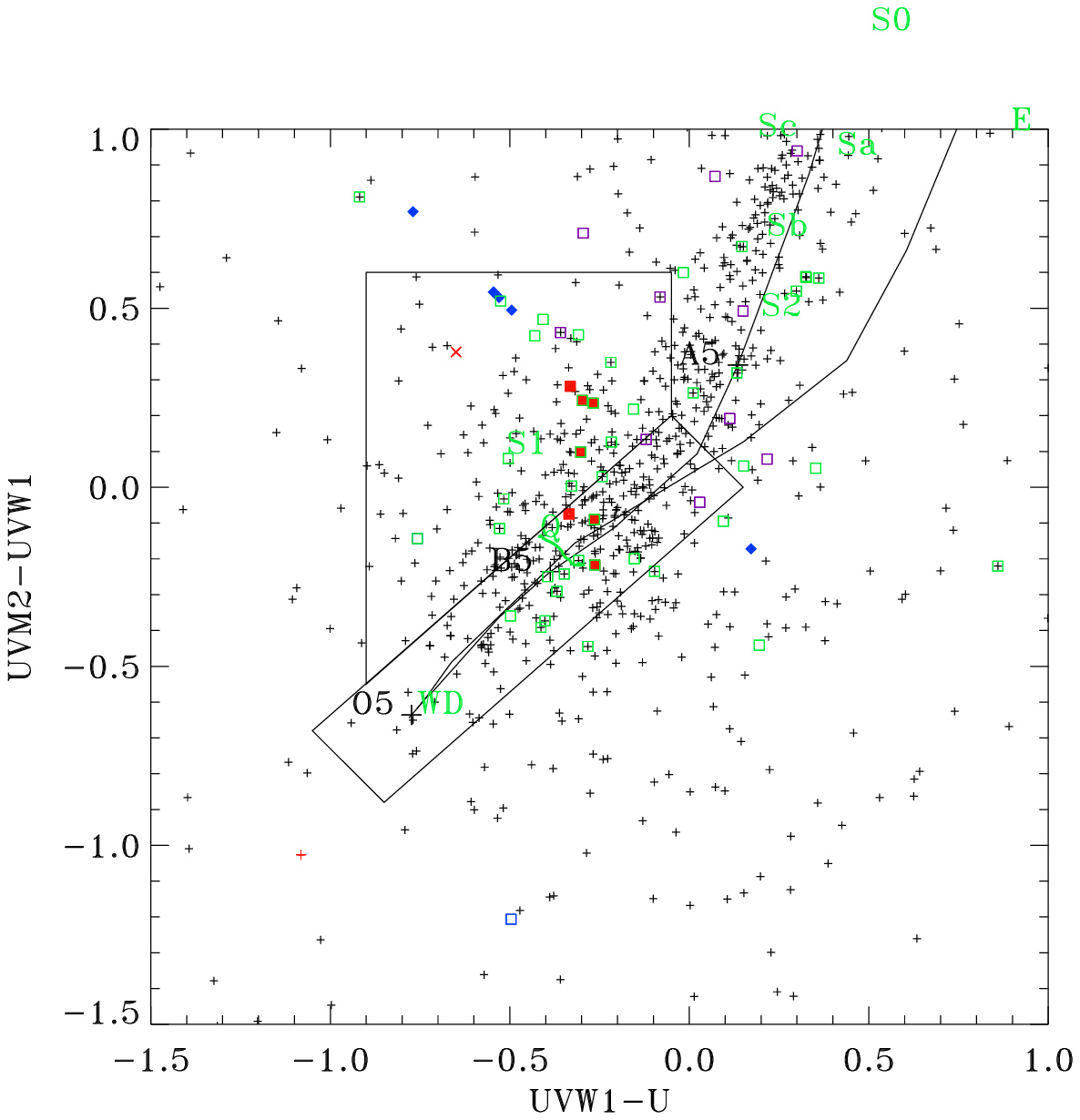}\plotone{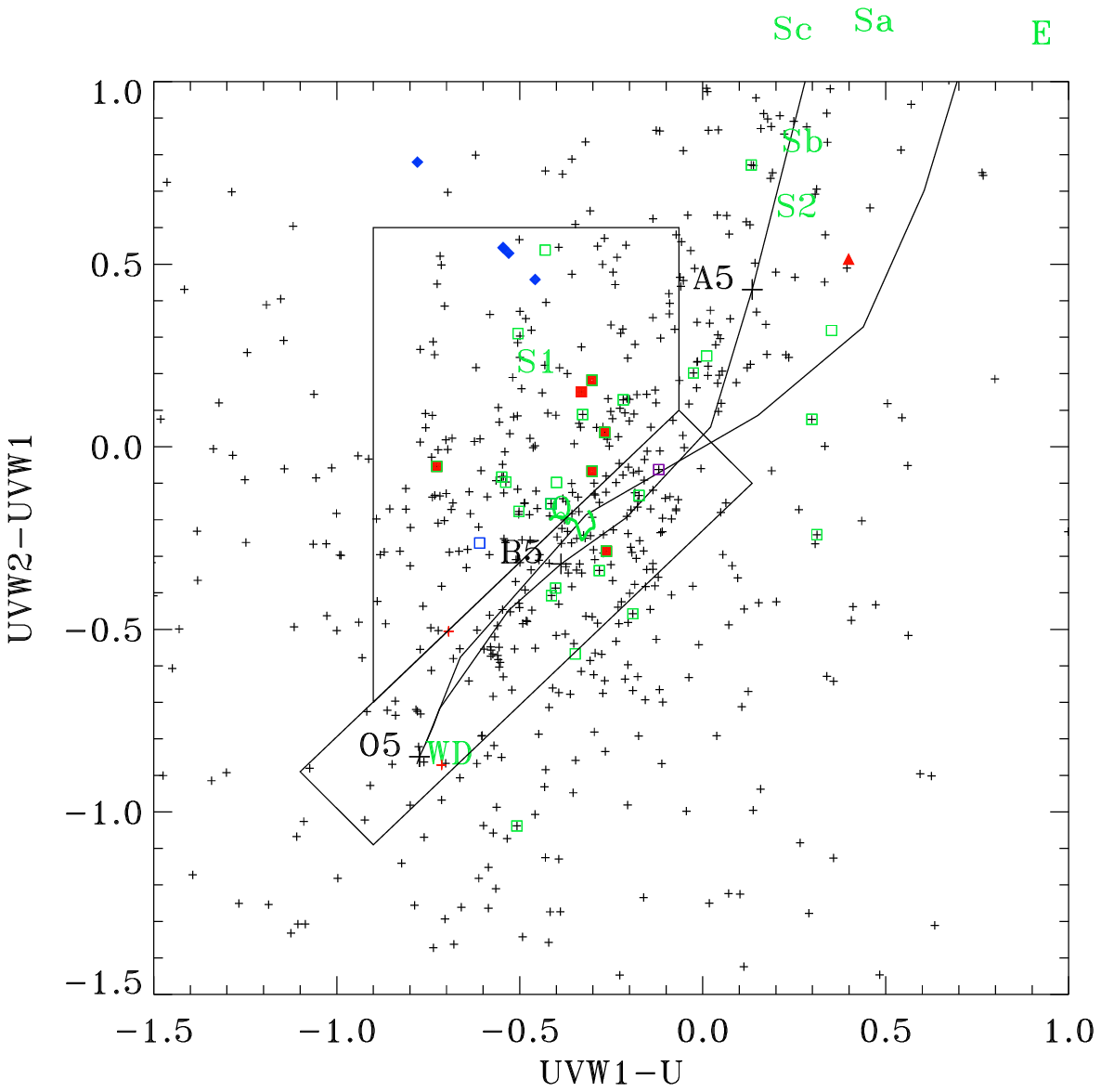}}
\centerline{\plotone{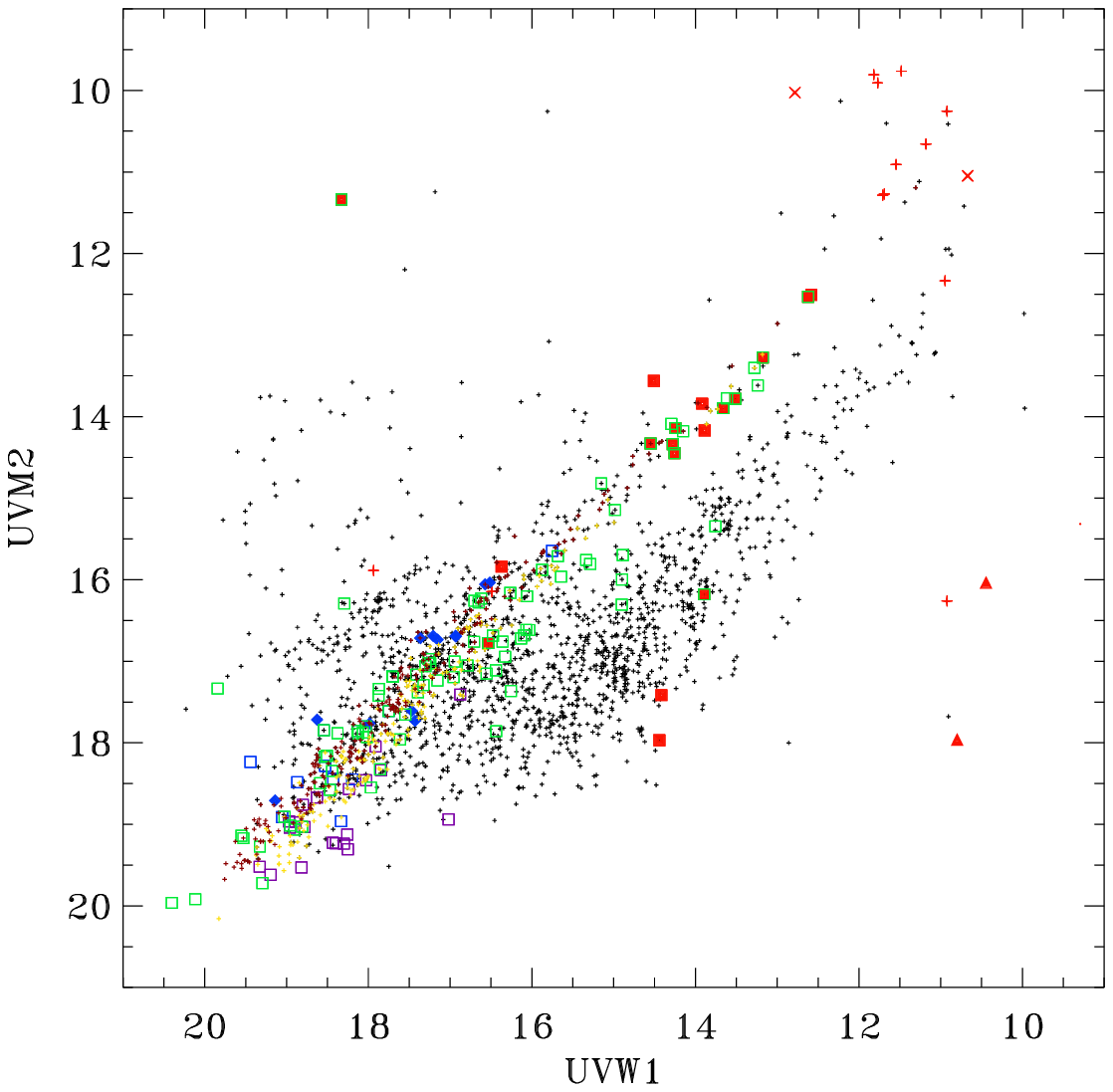}\plotone{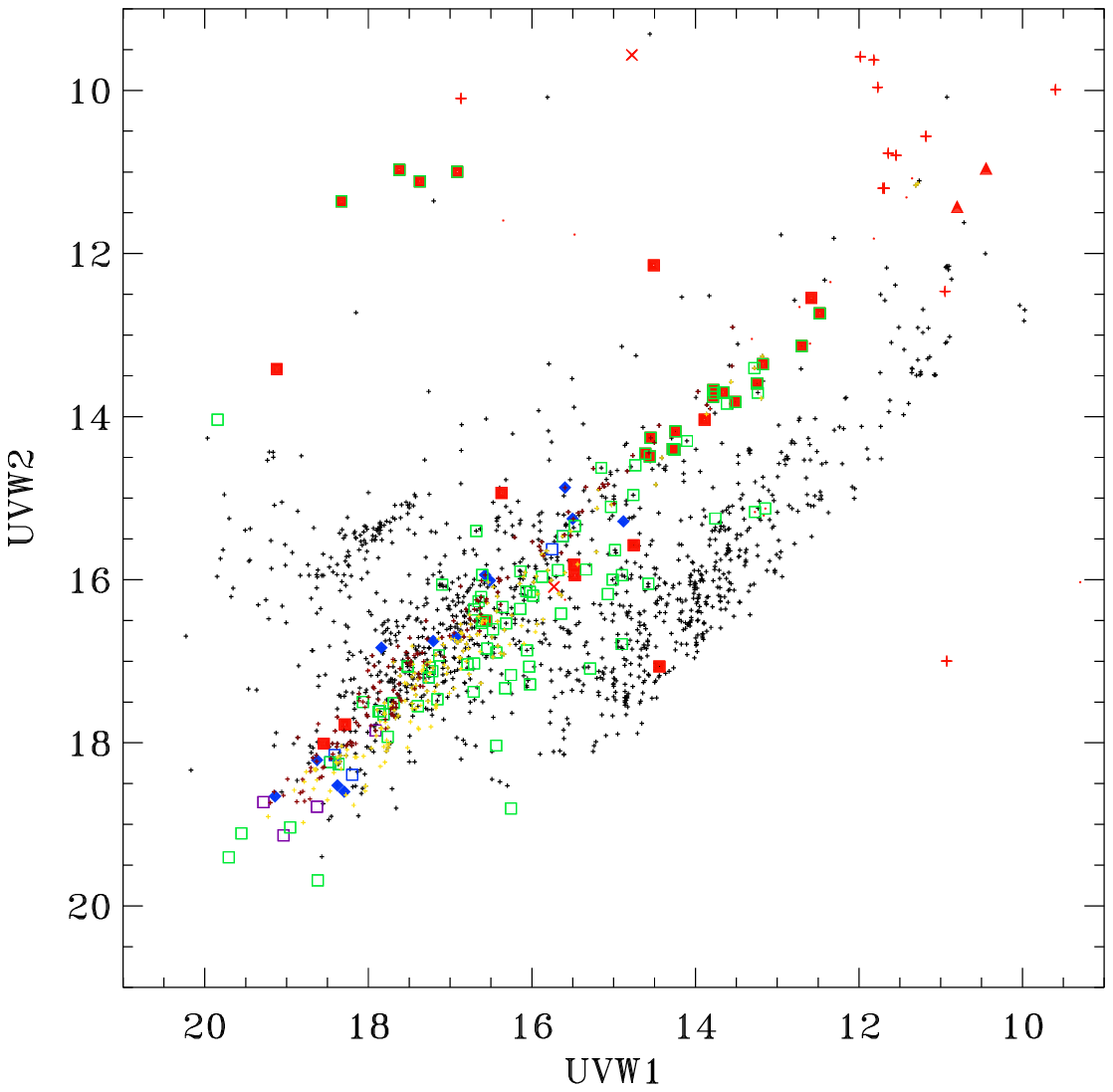}}
\caption{\footnotesize {\bf Top Left: }
The UVW1$-$U {\it versus} UVW2$-$UVW1 color-color diagram,
for all sources with $|b|>46\arcdeg$ and
with detection significances greater than three
and uncertainties $<0.1$ mag in each band.
{\it Red} symbols are sources observed by {\it FUSE}.
Boxes are AGN, {$+$} are main sequence and giant stars,
{$\times$} are PN stars, Diamonds are white dwarfs,
and Triangles are symbiotic stars.
The source types are taken from the {\it FUSE} master catalogue.
{\it Green} symbols are AGN from the 12$^{th}$ Veron catalogue.
{\it Blue} symbols are from the Sloan survey.
{\it Purple} symbols are from the Kauffman/Sloan catalogue of AGN.
The large boxes are the regions of interest for AGN (trapezoidal)
and upper main sequence stars (rectangular).
{\bf Top Right: }
The same plot for the UVW1$-$U {\it versus} UVW2$-$UVW1 color-color diagram.
{\bf Bottom Left: }
The UVW1 {\it versus} UVM2 diagram for the same sources.
Note that there are three ``sequences'' of sources,
the main one which runs almost along the diagonal
which is composed of early type stars and AGN,
a more diffuse band offset to the right
which is composed of late type stars with UVM2-UVW1$>1.5$,
and a very short one to the left
which is composed sources with UVM2-UVW1$<-2.0$;
these are almost entirely spurious sources
at the edge of the photocathode FOV.
{\bf Bottom Right: }
The UVW1 {\it versus} UVW2 diagram for the sources
in the UVW1$-$U {\it versus} UVW2$-$UVW1 color-color diagram.
The same three sequences are distinguishable.
\label{fig:fuse}}
\end{figure*}

The top panels of Figure~\ref{fig:fuse} show the OM color-color diagram
with the {\it FUSE} matches marked in red and coded by source type.
AGN found from cross-correlating the OMCat 
with a number of AGN catalogues have also been plotted.
The bulk of the {\it FUSE} observations with matching OM sources are AGN.
The difficulty is that the AGN, early type stars, 
extragalactic star-forming regions, and some galaxies 
overlap in these color-color diagrams.
We can define a long rectangular region 
along the early type star track
as being contaminated by stars.
Above and to the left is a large irregular region
that has a significant population of AGN.
The lower panels of Figure~\ref{fig:fuse} 
show the OM UVW1 and UVM2 (or UVW2) magnitudes of the sources.
Again, the OM sources matching {\it FUSE} observations 
have been marked in red and coded by {\it FUSE} type.
{\it FUSE} observed sources down to UVW1 magnitudes $\sim18.5$,
which suggests sources that are well detected by the OM
should have been observable by {\it FUSE}.

We have selected potential spectroscopic targets with $|b|>45\arcdeg$,
detection significance $>3$, $\sigma_U<0.1$ magnitudes,
$\sigma_{UVW1}<0.1$ magnitudes, 
and $\sigma_{UVM2}<0.1$ magnitudes or $\sigma_{UVW2}<0.1$ magnitudes,
which yielded $\sim$1500 UVM2 candidates and $\sim$1000 UVW2 candidates.
(Even with identical coverage, one would still expect fewer UVW2
candidates simply because of the lower response in the UVW2 filter.)
We then required the sources to fall within the boxes
shown in the color-color diagrams of Figure~\ref{fig:fuse},
reducing the number of candidates to $\sim$600.
We kept the candidates in the ``stellar'' box separate
from those in the more extended ``AGN'' box.
By visual inspection,
we then removed all sources that were actually optical defects
(bad pixels, ghosts, diffraction spikes, etc.)
or were emission regions in nearby galaxies,
or were diffuse with no point-like source.
This selection reduced the candidate list to $\sim$250.
Since we are interested in UV sources
not in previous catalogues we then removed sources
that had been observed by {\it FUSE}.

Combining the candidates obtained from the UVW1$-$U/UVM2$-$UVW1
color-color diagram with those obtained from the UVW1$-$U/UVW2$-$UVW1
color-color diagram, 
we found a total of 54 unique sources in the ``AGN'' box
and a total of 97 unique sources in the ``stellar'' box.
The magnitude distribution is similar to that of the survey as a whole;
of the ``AGN'' box sources, 5 are fainter than UVW1$=$18.5
and of the ``stellar'' box sources, 10 are fainter than UVW1$=$18.5.
The brightest 10\% of the sources in both boxes
are previously known AGN which were the explicit subject of study
for the corresponding X-ray observations.
Combining the two boxes,
we obtained $\sim$100 sources with UVW1$<$18.5 
that have not been previously studied.

There is still the matter of source classification
and the contamination of the source candidates 
(hoped to be stars and AGN) by galaxies.
Of the candidate sources, somewhat over a third
have SDSS counterparts with spectroscopic source types.
Of the sources in the ``AGN'' box, $\sim$80\% are categorized as QSO
and the remainder are categorized as galaxies.
Of the sources in the ``stellar'' box,
$\sim$40\% are categorized as QSO,
$\sim$40\% are categorized as galaxies,
and the remainder are categorized as stars.
Thus, while the ``AGN'' box is a much cleaner sample of AGN,
the ``stellar'' box still produces a similar number of AGN.

Of the candidates, only 9 (3 from the ``AGN'' box,
6 from the ``stellar'' box)
fall on or near high velocity clouds.
The remainder of the sources are uniformly
distributed across the high Galactic latitude sky
and will be useful for the study of the structure 
of the Galactic halo in the manner of \citet{wakker}.

\begin{figure}[h!]
\epsscale{1.0}
\plotone{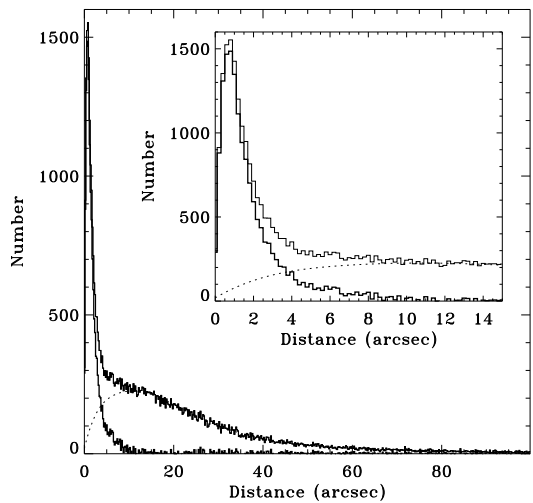}
\caption{\footnotesize {\it Light Histogram: } The distribution of the distance between
an X-ray source and the closest OM source.
{\it Dotted Curve: } The fitted probability distribution
for the distance between the X-ray sources and the closest OM source
for X-ray sources not correlated with OM sources (see text).
{\it Heavy Histogram: } The difference between those two distributions,
which should be the distribution of true matches.
\label{fig:xo_dist_stat}}
\end{figure}

\subsection{X-ray Counterparts}

\subsubsection{Matching the Catalogues}

We matched the {\it XMM-Newton} Serendipitous Source Catalogue (SSC,
The Second XMM-Newton Serendipitous Source Pre-release Catalogue, 
{\it XMM-Newton} Survey Science Centre, 2006) against the OMCat.
We first filtered out all of the \xmm\ serendipitous sources
that did not fall within the field of view of the corresponding OM observation.
Because we did not know the relative positional uncertainty to expect
when comparing the X-ray detections with the OM detections,
we matched the SSC to the OMCat in the same way that
we matched the USNO catalogue to the OMCat.
The distribution of the distance between the X-ray sources
and the closest OM source is shown in Figure~\ref{fig:xo_dist_stat};
the sharp peak at $\sim0.5\arcsec$ 
is presumably due to true matches between X-ray and OM sources 
while the broader peak is due to uncorrelated sources.
In order to determine the extent to which our matches
are contaminated by serendipitous alignments,
we calculated the probability distribution of Equation 1
for each source in the SSC that fell within the OM FOV
using the density of OM sources in the FOV
for the observation containing the SSC source.
Because some of the SSC sources do have real matches,
this calculation overestimates the probability distribution
of serendipitous matches.
In order to correct for this overestimation,
we assumed that sources with distances of $r<8\arcsec$
had a non-negligible probability of being real matches.
We then summed over the probability distribution
for $r>8\arcsec$ and normalized that to the total number
of matches found with $r>8\arcsec$.
We then used this normalized model distribution to
determine the fraction of matches at distance $r$
which are due to serendipitous coincidences of uncorrelated sources.

We have taken as ``real'' X-ray-OMCat matches all those 
with distances $r<2\farcs5$.
At $r=2\farcs5$ there are still four times as many real matches
as there are random coincidences 
but the contamination rate for $r<2\farcs5$ is $\sim10$\%.
Of the 53848 sources from the XMM Serendipitous Source catalogue
falling within the FOV of the corresponding OM observations
and having coordinate corrections from the USNO,
12986 have OM source counterparts, of which 1092 are expected to be spurious.
It should be noted that in matching the X-ray sources to the OM sources,
the USNO corrected coordinates provide a much closer match
than do the original OM coordinates.

\subsubsection{Results}

\begin{figure}[h!]
\epsscale{1.0}
\plotone{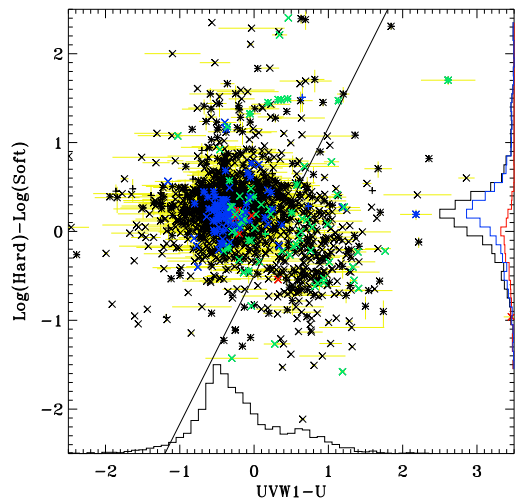}
\caption{\footnotesize The UV color plotted against
the log$_{10}$ of the 2-12/0.5-2 keV X-ray hardness ratio
for all sources with good UV or X-ray colors and $|b|>30\arcdeg$.
For clarity, error bars are shown only
when the Hard/Soft hardness ratio has a significance greater than 3
or the UVW1$-$U uncertainty is $<0.477$ magnitudes.
The diagonal line separates the sources clustered around the ``AGN'' colors
from those clustered around the ``stellar'' colors.
Histograms of the projected distributions are shown along each axis.
Along the vertical axis the distribution of the X-ray hardness ratio is plotted
for all of the sources (black),
the sources to the upper left of the line (blue),
and to the lower right of the line (red).
Along the horizontal axis is plotted the UV color ratios for all the sources.
The {\bf blue} symbols are SDSS sources classified as QSO,
the {\bf green} symbols are SDSS sources classified as galaxies,
and the {\bf red} symbols are SDSS sources classified as stars.
Since the SDSS sources chosen for spectroscopy were selected
to have a low probability of being stars,
based on their colors and being unresolved, the low number of stars,
and that fact that they have AGN-like colors, should not be surprising.
The $\times$ indicate sources with good UV colors,
the $+$ indicate sources with good X-ray colors,
and source with both good UV and X-ray colors are marked with both symbols.
For clarity, error bars equivalent to S/N$<3$ are not shown.
\label{fig:color_hard}}
\end{figure}

The bulk of the X-ray sources are expected to be background AGN 
and Galactic stars.
Given the observational interest in galaxies,
there will also be a small, probably negligible, contribution
from extragalactic X-ray binaries and star-forming regions.
The question of interest when comparing catalogues at different energies
is whether the X-ray sources detected in the UV fall in distinctive portions
of X-ray/UV color/hardness diagrams.
Figure~\ref{fig:color_hard} compares the UVW1$-$U color
(the UV color for which we have the greatest number of sources)
with the 2.0-12.0/0.5-2.0 keV X-ray hardness.
From Figure~\ref{fig:color_color}, 
we expect AGN to have UVW1$-$U$\sim0.4$ 
and stars to have UVW1$-$U$<0$.
The same behavior is observed in the color-hardness diagram,
with an additional separation in X-ray hardness;
the X-ray sources with UV colors of AGN
have Log(Hard)-Log(Soft) peaked around 0.2
and the X-ray sources with UV colors of stars
have a broad distribution peaking at Log(Hard)-Log(Soft)$<0$.
This distribution reflects the well understood difference between
the soft thermal X-ray spectra of stars
and the harder power-law X-ray spectra of AGN.
The combination of X-ray and UV colors are probably 
a more powerful star-AGN discriminator than either alone.
The diagonal line in Figure~\ref{fig:color_hard}
marks a reasonable separation between these two categories.

To test this separation we have cross-correlated the sources
with good X-ray and UV data with the SDSS DR5 catalogue of sources
with classifications made from the SDSS spectra.
Indeed, the AGN (blue) are clustered as expected.
The bulk of the galaxies (green) have stellar colors,
but many have AGN-like colors.
Since the SDSS sources chosen for spectroscopy were selected
to have a low probability of being stars, 
based on their colors and being unresolved, the low number of stars (red), 
and that fact that they have AGN-like colors, should not be surprising.

\begin{figure}
\epsscale{1.0}
\plotone{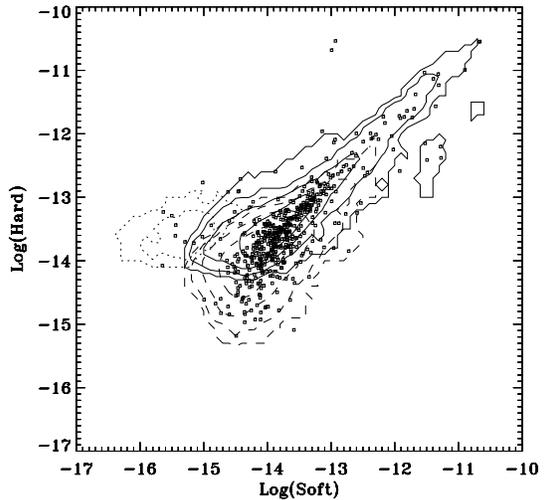}
\caption{\footnotesize The contours show the density of sources from the entire SSC
in the soft {\it versus} hard band parameter space.
The fluxes are in erg cm$^{-2}$ s$^{-1}$.
{\bf Solid:} sources with $3\sigma$ detections in both bands,
{\bf Dashed:} sources with $3\sigma$ detections in only the hard band, and
{\bf Dotted:} sources with $3\sigma$ detections in only the soft band.
The points are sources with UVW1$-$U colors.
\label{fig:hard_uv}}
\end{figure}

The contours in Figure~\ref{fig:hard_uv} 
show the density of SSC catalogue sources
in the Hard (2.0-12.0 keV) {\it versus} Soft (0.5-2.0 keV) band space;
the points show the locations of sources with UVW1$-$U colors.
Not surprisingly, 
many of the sources with only soft X-ray detections have UV detections
while there are only a few sources that have
only hard X-ray detections that also have UV detections.
The hard X-ray sources with UV detections tend to have AGN-like colors.
To state it in the converse, very few hard X-ray sources
(i.e., strongly absorbed sources) have UV counterparts.

\begin{figure}[]
\epsscale{1.0}
\plotone{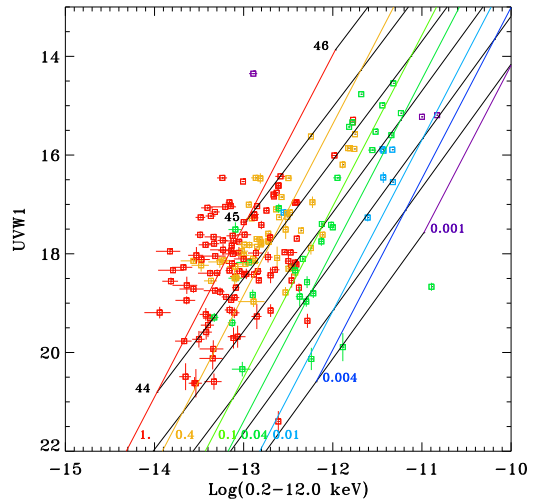}
\caption{\footnotesize SSC sources with OM counterparts in the UVW1 band
and AGN counterparts in the SDSS spectroscopic survey.
{\bf Small boxes:} sources are color-coded by redshift:
$0\le z<0.04$: dark blue,
$0.04\le z<0.1$: light blue,
$0.1\le z<0.4$: green,
$0.4\le z<1.0$: orange,
$1.0\le z$: red.
{\bf Solid lines:} the expected tracks for QSOs,
color coded by redshift:
$z=0.001$: purple,
$z=0.004$: dark blue,
$z=0.01$: light blue,
$z=0.04$: dark green,
$z=0.1$: light green,
$z=0.4$: yellow,
$z=1.0$: red.
The black lines are lines of constant rest-frame
0.2-12.0 keV band luminosity from $\log{L_x}=40$
to $\log{L_x}=46$.
\label{fig:sdss_agn}}
\end{figure}

The SSC sources with UV counterparts that also have QSO counterparts 
in the SDSS spectroscopic survey are shown in Figure~\ref{fig:sdss_agn}.
The sources cluster along the tracks expected 
if the X-ray and UV flux are related by
\begin{equation}
\log_{10}{L_{2 keV}}=0.721\log_{10}{L_{2500\AA}}+4.531
\end{equation}
as found by \citet{ssbaklsv2006},
where the monochromatic luminosities are in erg s$^{-1}$ Hz$^{-1}$.
We assumed the QSO template spectrum from \citet{kea1996}
for calculating the UV flux and magnitudes.
We have assumed the photon index of the X-ray spectrum to be $\Gamma=2.0$,
and (for the plotted tracks) no internal absorption.
Assuming internal absorption moves the tracks by $\log{f_x}\sim0.1$,
so the effect is not strong for this X-ray energy band.
Although there is significantly more scatter than one would expect
from the uncertainties, the sources do lie roughly where expected.
It should be noted that this small sample of sources
(\xmm\ X-ray sources with OM counterparts 
and SDSS spectroscopic survey counterparts)
seems to be a relatively unbiased sample of the SDSS spectral catalogue;
the distribution of spectroscopic classification and redshift in the sample
matches the distribution of all of the sources in 
the $12\arcmin$ radius SDSS spectroscopic catalogue extracts
centered on the OM pointing directions.
With the larger sample of OM sources with X-ray counterparts
one will be able to begin to address the source of this scatter.

\section{Summary}
The OMCat provides a quick source of photometric data of point-like sources
in the optical and near ultraviolet 
over an increasing fraction of the sky.
The OMCat will continue to be augmented at the HEASARC
as the \xmm\ data becomes public.
Although the short average exposure places a relatively high detection limit
compared to optical catalogues such as USNO-B,
the current detection limit provides suitably bright targets
for current UV spectrometers,
and provides high angular resolution data suitable 
for meaningful comparison with \galex\ images.
A first glance at the AGN counterparts in the OMCat 
suggests their UV properties are not particularly unexpected.
No doubt more targeted querying of the OMCat
will produce interesting science in several different fields.

It is expected that the current version of the OMCat
will be superseded by a sigificantly improved processing
produced by the \xmm\ project.
Until that catalogue is produced, 
the HEASARC will continue to augment the current version.

\acknowledgements
We would like to thank the referee, Marcel Ag{\"u}eros,
for many useful comments.
We would like to thank Martin Still for many useful discussions
about the current and future OMCats.
We would like to thank Mike Arida for his help in making
the OM data available through the HEASARC.
We would like to thank Karen Levay, Randy Thompson, and Rick White
for their help in making this data MAST accessible.
We would like to thank Luciana Bianchi for many useful discussions,
as well as B.G. Anderson for his help with {\it FUSE}.
We would also like to thank Lorenzo Principe
for his help with this work.

This work would not have been possible without SAS,
the Science Survey Consortium, 
and the OM PI team led by K. Mason at the Mullard Space Science Laboratory.

This research used data obtained through the Browse facility
of the High Energy Astrophysics Science Archive Research Center (HEASARC). 
The OM object catalogue is now available through Browse
\footnote{http://heasarc.gsfc.nasa.gov/docs/archive.html}
and in a static ASCII version\footnote{
http://heasarc.gsfc.nasa.gov/FTP/heasarc/dbase/tdat\_files/ heasarc\_xmmomcat.tdat.gz}.
Several databases accessed through browse are actually held by 
the {\it VizieR} service at http://vizier.u-strasbg.fr.
\citep{vizier}.

This research has made use of data from the \xmm\ 
Serendipitous Source Catalogue,
a collaborative project of the \xmm\ Survey Science Center Consortium,
http://xmmssc-www.star.le.ac.uk.

This research has made use of the \galex\ GR2/GR3 database/archive
at the Multimission Archive at STScI (MAST).
STScI is operated by the Association of Universities for Research in Astronomy, 
Inc., under NASA contract NAS5-26555. 
Support for MAST for non-HST data is provided by the NASA 
Office of Space Science via grant NAG5-7584 and by other grants and contracts.

This research has made use of data obtained from Data Release 5 
of the Sloan Digital Sky Survey (SDSS).
Funding for the Sloan Digital Sky Survey has been provided by 
the Alfred P. Sloan Foundation, the Participating Institutions, 
the National Aeronautics and Space Administration, the National Science Foundation,
the U.S. Department of Energy, the Japanese Monbukagakusho, 
and the Max Planck Society. The SDSS Web site is http://www.sdss.org/. 
The SDSS is managed by the Astrophysical Research Consortium (ARC) 
for the Participating Institutions. 
The Participating Institutions are The University of Chicago, 
Fermilab, the Institute for Advanced Study, the Japan Participation Group, 
The Johns Hopkins University, Los Alamos National Laboratory, 
the Max-Planck-Institute for Astronomy (MPIA), 
the Max-Planck-Institute for Astrophysics (MPA), New Mexico State University, 
University of Pittsburgh, Princeton University, 
the United States Naval Observatory, and the University of Washington.


\end{document}